\documentclass[aos,MSNbibl,seceqn,nameyear,dvips]{arximspdf}
\usepackage{graphicx}
\doi{10.1214/13-AOS1140} 
\volume{41}
\issue{5}
\pubyear{2013}
\firstpage{2263}
\lastpage{2291}

\makeatletter

\newcommand{\rrVert}{\Vert}
\newcommand{\rrvert}{\vert}
\newcommand{\llVert}{\Vert}
\newcommand{\llvert}{\vert}
\def\implies{\Rightarrow}
\newtheorem{thmm}{Theorem}
\newproclaim{defn}[thmm]{Definition}
\newtheorem{prop}[thmm]{Proposition}
\newproclaim{rem}[thmm]{Remark}
\newtheorem{lem}[thmm]{Lemma}
\newproclaim{example}[thmm]{Example}
\newtheorem{cor}[thmm]{Corollary}
\newcommand{\bb}{\mathbb}
\newcommand{\Cal}{\mathcal}
\newcommand{\eqref}[1]{(\ref{#1})}
\makeatother

\begin{document}
\begin{frontmatter}

\title{Equivalence of distance-based and RKHS-based statistics in hypothesis testing}
\runtitle{Distance-based and RKHS-based statistics}

\begin{aug}
\author[a]{\fnms{Dino}~\snm{Sejdinovic}\corref{}\ead[label=e1]{dino.sejdinovic@gmail.com}},
\author[b]{\fnms{Bharath}~\snm{Sriperumbudur}\ead[label=e2]{b.sriperumbudur@statslab.cam.ac.uk}},
\author[a]{\fnms{Arthur}~\snm{Gretton}\ead[label=e4]{arthur.gretton@gmail.com}}
\and
\author[c]{\fnms{Kenji}~\snm{Fukumizu}\ead[label=e3]{fukumizu@ism.ac.jp}}
\runauthor{Sejdinovic, Sriperumbudur, Gretton and Fukumizu}
\affiliation{University College London,
University of Cambridge,
University College London and Max Planck Institute for Intelligent Systems,
T\"{u}bingen, and
Institute of Statistical Mathematics, Tokyo}

\address[a]{D. Sejdinovic\\
A. Gretton\\
Gatsby Computational Neuroscience Unit\\
University College London\\
17 Queen Square\\
London, WC1N 3AR\\
United Kingdom\\
\printead{e1}\\
\phantom{E-mail:\ }\printead*{e4}}
\address[b]{B. Sriperumbudur\\
Statistical Laboratory\\
Center for Mathematical Sciences\\
University of Cambridge\\
Wilberforce Road\\
Cambridge CB3 0WB\\
United Kingdom\\
\printead{e2}}
\address[c]{K. Fukumizu\\
Institute of Statistical Mathematics\\
10-3 Midoricho, Tachikawa\\
Tokyo 190-8562\\
Japan\\
\printead{e3}}
\end{aug}

\received{\smonth{7} \syear{2012}}
\revised{\smonth{3} \syear{2013}}

%
\begin{abstract}
We provide a unifying framework linking two
classes of statistics used in two-sample and independence testing: on
the one hand, the energy distances and distance covariances from the
statistics literature; on the other, maximum mean discrepancies (MMD),
that is, distances between embeddings of distributions to reproducing
kernel Hilbert spaces (RKHS), as established in machine learning. In
the case where the energy distance is computed with a semimetric of
negative type, a positive definite kernel, termed distance kernel, may
be defined such that the MMD corresponds exactly to the energy
distance. Conversely, for any positive definite kernel, we can
interpret the MMD as energy distance with respect to some negative-type
semimetric. This equivalence readily extends to distance covariance
using kernels on the product space. We determine the class of
probability distributions for which the test statistics are consistent
against all alternatives. Finally, we investigate the performance of
the family of distance kernels in two-sample and independence tests: we
show in particular that the energy distance most commonly employed in
statistics is just one member of a parametric family of kernels, and
that other choices from this family can yield more powerful tests.
\end{abstract}

%

\end{frontmatter}

\section{Introduction}
\label{submission}

The problem of testing statistical hypotheses in high dimensional
spaces is particularly challenging, and has been a recent focus of
considerable work in both the statistics and the machine learning
communities. On the statistical side, two-sample testing in Euclidean
spaces (of whether two independent samples are from the same distribution,
or from different distributions) can be accomplished using a so-called
energy distance as a statistic [\citeauthor{Szekely2004} (\citeyear{Szekely2004,Szekely2005}), \citet{Baringhaus2004}].
Such tests are consistent against all alternatives as long as the
random variables have finite first moments. A related dependence measure
between vectors of high dimension is the distance covariance [\citet
{Szekely2007,SzeRiz09}],
and the resulting test is again consistent for variables with bounded
first moment. The distance covariance has had a major impact in the
statistics community, with \citet{SzeRiz09} being accompanied by
an editorial introduction and discussion. A~particular advantage of
energy distance-based statistics is their compact representation in
terms of certain expectations of pairwise Euclidean distances, which
leads to straightforward empirical estimates. As a follow-up work,
\citet{Lyons2011} generalized the notion of distance covariance to
metric spaces of negative type (of which Euclidean spaces are a special
case).

On the machine learning side, two-sample tests have been formulated
based on embeddings of probability distributions into reproducing
kernel Hilbert spaces [\citeauthor{GreBorRasSchetal07c} (\citeyear{GreBorRasSchetal07c,Gretton2012})], using
as the test statistic the difference between these embeddings: this
statistic is called the maximum mean discrepancy (MMD). This distance
measure was also applied to the problem of testing for independence,
with the associated test statistic being the Hilbert--Schmidt independence
criterion (HSIC) [\citeauthor{GreBorRasSchetal07c} (\citeyear{GreBouSmoSch05,GreFukTeoSonetal08}), \citet{SmoGreSonSch07,Zhang2011}].
Both tests are shown to be consistent against all alternatives when
a characteristic RKHS is used [\citet{FukSriGreSch09,SriGreFukLanetal10}].

Despite their striking similarity, the link between energy distance-based
tests and kernel-based tests has been an open question. In the discussion
of [\citet{SzeRiz09}, \citet{GreFukSri09}, page~1289] first explored
this link in the context of independence testing, and found that interpreting
the distance-based independence statistic as a kernel statistic is
not straightforward, since Bochner's theorem does not apply to the
choice of weight function used in the definition of the distance covariance
(we briefly review this argument in Section~\ref{subCharacteristic-function-interpre}).
\citet{SzeRiz09}, Rejoinder, page~1303, confirmed that the link between
RKHS-based dependence measures and the distance covariance remained
to be established, because the weight function is not integrable.
Our contribution resolves this question, and shows that RKHS-based
dependence measures are precisely the formal extensions of the distance
covariance, where the problem of nonintegrability of weight functions
is circumvented by using translation-variant kernels, that is, \emph
{distance-induced
kernels}, introduced in Section~\ref{subDistance-kernels}.\looseness=-1

In the case of two-sample testing, we demonstrate that energy distances
are in fact maximum mean discrepancies arising from the same family
of distance-induced kernels. A number of interesting consequences
arise from this insight: first, as the energy distance (and distance
covariance) derives from a particular choice of a kernel, we can consider
analogous quantities arising from other kernels, and yielding more
sensitive tests. Second, in relation to \citet{Lyons2011}, we obtain
a new family of characteristic kernels arising from general semimetric
spaces of negative type, which are quite unlike the characteristic
kernels defined via Bochner's theorem [\citet{SriGreFukLanetal10}].
Third, results from [\citet{GreFukHarSri09,Zhang2011}] may be applied
to obtain consistent two-sample and independence tests for the energy
distance, without using bootstrap, which perform much better than
the upper bound proposed by \citet{Szekely2007} as an alternative
to the bootstrap.

In addition to the energy distance and maximum mean discrepancy, there
are other well-known discrepancy measures between two probability
distributions, such as the Kullback--Leibler divergence, Hellinger
distance and total variation distance, which belong to the class of
$f$-divergences. Another popular family of distance measures on probabilities
is the integral probability metric [\citet{Mueller97}], examples of
which include the Wasserstein distance, Dudley metric and Fortet--Mourier
metric. \citet{SriFukGreSchetal11} showed that MMD is an integral
probability metric and so is energy distance, owing to the equality
(between energy distance and MMD) that we establish in this paper.
On the other hand, \citet{SriFukGreSchetal11} also showed that MMD
(and therefore the energy distance) is not an $f$-divergence, by
establishing the total variation distance as the only discrepancy
measure that is both an IPM and $f$-divergence.

The equivalence established in this paper has two major implications
for practitioners using the energy distance or distance covariance
as test statistics. First, it shows that these quantities are members
of a much broader class of statistics, and that by choosing an alternative
semimetric/kernel to define a statistic from this larger family, one
may obtain a more sensitive test than by using distances alone. Second,
it shows that the principles of energy distance and distance covariance
are readily generalized to random variables that take values in general
topological spaces. Indeed, kernel tests are readily applied to structured
and non-Euclidean domains, such as text strings, graphs and groups
[\citet{FukSriGreSch09}].

The structure of the paper is as follows: in Section~\ref{secDistance-based-approach},
we introduce semimetrics of negative type, and extend the notions
of energy distance and distance covariance to semimetric spaces of
negative type. In Section~\ref{seckernel-based-approach}, we provide
the necessary definitions from RKHS theory and give a review of the
maximum mean discrepancy (MMD) and the Hilbert--Schmidt independence
criterion (HSIC), the RKHS-based statistics used for two-sample and
independence testing, respectively. In Section~\ref{secCorrespondencePDKvsNTsM},
the correspondence between positive definite kernels and semimetrics
of negative type is developed, and it is applied in Section~\ref{secEquivalences}
to show the equivalence between a (generalized) energy distance and
MMD (Section~\ref{subenergydistancewithkernels}), as well as
between a (generalized) distance covariance and HSIC (Section~\ref{subdcovwithkernels}).
We give conditions for these quantities to\vadjust{\goodbreak} distinguish between probability
measures in Section~\ref{secDistinguishing-probability-distributions},
thus obtaining a new family of characteristic kernels. Empirical estimates
of these quantities and associated two-sample and independence tests
are described in Section~\ref{secConsistency}. Finally, in Section~\ref{secExperiments}, we investigate the performance of the test
statistics on a variety of testing problems.

This paper extends the conference publication [\citet{Sejdinovic2012}],
and gives a detailed technical discussion and proofs which were omitted
in that work.

\section{Distance-based approach}\label{secDistance-based-approach}

This section reviews the distance-based approach to two-sample and
independence testing, in its general form. The generalized energy
distance and distance covariance are defined.

\subsection{Semimetrics of negative type}\label{subSemimetrics-negative-type}

We will work with the notion of a semimetric of negative type on a
nonempty set $\mathcal{Z}$, where the ``distance'' function need
not satisfy the triangle inequality. Note that this notion of semimetric
is different to that which arises from the seminorm (also called the
pseudonorm), where the distance between two distinct points can be
zero.
%
\begin{defn}[(Semimetric)]
 Let $\mathcal{Z}$ be a nonempty set and let
$\rho\dvtx \mathcal{Z}\times\mathcal{Z}\to[0,\infty)$ be a function such
that $\forall z,z'\in\mathcal{Z}$,
\begin{longlist}[1.]
\item[1.]$\rho(z,z')=0$ if and only if $z=z'$, and
\item[2.]$\rho(z,z')=\rho(z',z)$.
\end{longlist}

Then $(\mathcal{Z},\rho)$ is said to be a semimetric space and $\rho$
is called a semimetric on $\mathcal{Z}$.
\end{defn}
%

%
\begin{defn}[(Negative type)] The semimetric space $(\mathcal{Z},\rho)$
is said to have negative type if $\forall n\geq2$, $z_{1},\ldots
,z_{n}\in\mathcal{Z}$,
and $\alpha_{1},\ldots,\alpha_{n}\in\mathbb{R}$, with\break $\sum_{i=1}^{n}\alpha_{i}=0$,
%
%
\begin{equation}
\sum_{i=1}^{n}\sum
_{j=1}^{n}\alpha_{i}\alpha_{j}
\rho (z_{i},z_{j})\leq 0.\label{eqCND}
\end{equation}
Note that in the terminology of \citet{BerChrRes84}, $\rho$ satisfying
\eqref{eqCND} is said to be a \emph{negative definite} function.
The following proposition is derived from \citet{BerChrRes84}, Corollary 2.10, page 78,
and Proposition 3.2, page 82.
\end{defn}
%

%
\begin{prop}
$ $\label{prosemimetrichilbertianmetric}
\begin{longlist}[1.]
\item[1.] If $\rho$ satisfies \eqref{eqCND}, then so does $\rho^{q}$, for
$0<q<1$.
\item[2.]$\rho$ is a semimetric of negative type if and only if there exists
a Hilbert space $\mathcal{H}$ and an injective map $\varphi\dvtx \mathcal
{Z}\to\mathcal{H}$,
such that
%
%
\begin{equation}
\rho\bigl(z,z'\bigr)=\bigl\llVert \varphi(z)-\varphi
\bigl(z'\bigr)\bigr\rrVert _{\mathcal
{H}}^{2}.\label{eqrhoviaHilbert}
\end{equation}
\end{longlist}
\end{prop}\eject
The second part of the proposition shows that $(\mathbb{R}^{d},\llVert \cdot-\cdot\rrVert ^{2})$
is of negative type, and by taking $q=1/2$ in the first part, we
conclude that all Euclidean spaces are of negative type. In addition,
whenever $\rho$ is a semimetric of negative type, $\rho^{1/2}$ is
a metric of negative type, that is, even though $\rho$ may not satisfy
the triangle inequality, its square root must do if it obeys \eqref{eqCND}.

\subsection{Energy distance}

Unless stated otherwise, we will assume that $\mathcal{Z}$ is any
topological space on which Borel measures can be defined. We will
denote by $\mathcal{M}(\mathcal{Z})$ the set of all finite signed
Borel measures on $\mathcal{Z}$, and by $\mathcal{M}_{+}^{1}(\mathcal{Z})$
the set of all Borel probability measures on $\mathcal{Z}$.

The energy distance was introduced by \citeauthor{Szekely2004}
(\citeyear{Szekely2004,Szekely2005})
and independently by \citet{Baringhaus2004} as a measure of statistical
distance between two probability measures $P$ and $Q$ on $\mathbb{R}^{d}$
\emph{with finite first moments}, given by
%
%
\begin{equation}
\qquad D_{E}(P,Q)=2\mathbb{E}_{ZW}\llVert Z-W\rrVert
_{2}-\mathbb {E}_{ZZ'}\bigl\llVert Z-Z'\bigr
\rrVert _{2}-\mathbb{E}_{WW'}\bigl\llVert W-W'
\bigr\rrVert _{2},\label{eqenergydistance}
\end{equation}
where $Z,Z'\stackrel{\mathrm{i.i.d.}}{\sim}P$ and $W,W'\stackrel{\mathrm{i.i.d.}}{\sim}Q$.
The moment condition is required to ensure that the expectations in
\eqref{eqenergydistance} is finite. $D_{E}(P,Q)$ is always nonnegative,
and is strictly positive if $P\neq Q$. In scalar case, it coincides
with twice the Cram\'{e}r--Von Mises distance.

Following \citet{Lyons2011}, the notion can be generalized to a metric
space of negative type, which we further extend to semimetrics. Before
we proceed, we need to first introduce a moment condition w.r.t. a
semimetric $\rho$.
%
\begin{defn}
For $\theta>0$, we say that $\nu\in\mathcal{M}(\mathcal{Z})$ has
a finite $\theta$-moment with respect to a semimetric $\rho$ of
negative type if there exists $z_{0}\in\mathcal{Z}$, such that $\int
\rho
^{\theta}(z,z_{0}) \,d\vert\nu\vert(z)<\infty$.
We denote
%
%
\begin{equation}
\mathcal{ M}_{\rho}^{\theta}(\mathcal{Z})= \biggl\{ \nu\in\mathcal
{M}(\mathcal{Z}) \dvtx \exists z_{0}\in\mathcal{Z}\mbox{ s.t. }\int\rho
^{\theta
}(z,z_{0}) \,d\llvert \nu\rrvert (z)<\infty \biggr\}.
\label{mrho}
\end{equation}
\end{defn}

We are now ready to introduce a general energy distance $D_{E,\rho}$.
%
\begin{defn}
\label{defenergydistance}Let $(\mathcal{Z},\rho)$ be a semimetric
space of negative type, and let $P,Q\in\mathcal{M}_{+}^{1}(\mathcal
{Z})\cap\mathcal{M}_{\rho}^{1}(\mathcal{Z})$.
The energy distance between $P$ and $Q$, w.r.t. $\rho$ is
%
%
\begin{equation}
D_{E,\rho}(P,Q)=2\mathbb{E}_{ZW}\rho(Z,W)-\mathbb{E}_{ZZ'}
\rho \bigl(Z,Z'\bigr)-\mathbb{E}_{WW'}\rho
\bigl(W,W'\bigr),\label{eqenergydistancegen}
\end{equation}
where $Z,Z'\stackrel{\mathrm{i.i.d.}}{\sim}P$ and $W,W'\stackrel{\mathrm{i.i.d.}}{\sim}Q$.
\end{defn}
If $\rho$ is a metric, as in [\citet{Lyons2011}], the moment condition
$P,Q\in\mathcal{M}_{\rho}^{1}(\mathcal{Z})$ is easily seen to be
sufficient for the existence of the expectations in \eqref{eqenergydistancegen}.
Namely, if we take $z_{0},w_{0}\in\mathcal{Z}$ such that $\mathbb
{E}_{Z}\rho(Z,z_{0})<\infty$,
$\mathbb{E}_{W}\rho(W,w_{0})<\infty$, then the triangle inequality
implies:
\[
\mathbb{E}_{ZW}\rho(Z,W)  \leq \mathbb{E}_{Z}\rho
(Z,z_{0})+\mathbb {E}_{W}\rho(W,w_{0})+
\rho(z_{0},w_{0})<\infty.
\]

If $\rho$ is a general semimetric, however, a different line of reasoning
is needed, and we will come back to this condition in Remark \ref
{remark-sufficiency-of-moment-condition},
where its sufficiency will become clear using the link between positive
definite kernels and negative-type semimetrics established in Section~\ref{secCorrespondencePDKvsNTsM}.

Note that the energy distance can equivalently be represented in the
integral form,
%
%
\begin{equation}
D_{E,\rho}(P,Q)  =  -\int\rho\, d \bigl([P-Q]\times[P-Q] \bigr),\label{eqintegralexpressionenergydistance}
\end{equation}
whereby the negative type of $\rho$ implies the nonnegativity of
$D_{E,\rho}$, as discussed by \citeauthor{Lyons2011} [(\citeyear{Lyons2011}), page 10].

\subsection{Distance covariance}

A related notion to the energy distance is that of distance covariance,
which measures dependence between random variables. Let $X$ be a
random vector on $\mathbb{R}^{p}$ and $Y$ a random vector on $\mathbb{R}^{q}$.
The distance covariance was introduced by \citet{Szekely2007,SzeRiz09}
to address the problem of testing and measuring dependence between
$X$ and $Y$ in terms of a weighted $L_{2}$-distance between characteristic
functions of the joint distribution of $X$ and $Y$ and the product
of their marginals. As a particular choice of weight function is used
(we discuss this further in Section~\ref{subCharacteristic-function-interpre}),
it can be computed in terms of certain expectations of pairwise Euclidean
distances,
%
%
\begin{eqnarray}\label{eqdCovintermsofdistances}
\mathcal{V}^{2}(X,Y) & = & \mathbb{E}_{XY}
\mathbb{E}_{X'Y'}\bigl\llVert X-X'\bigr\rrVert _{2}
\bigl\llVert Y-Y'\bigr\rrVert _{2}
\nonumber\\
& &{} + \mathbb{E}_{X}\mathbb{E}_{X'}\bigl\llVert
X-X'\bigr\rrVert _{2}\mathbb{E}_{Y}
\mathbb{E}_{Y'}\bigl\llVert Y-Y'\bigr\rrVert _{2}
\\
& &{} - 2\mathbb{E}_{XY} \bigl[\mathbb{E}_{X'}\bigl\llVert
X-X'\bigr\rrVert _{2}\mathbb{E}_{Y'}\bigl\llVert
Y-Y'\bigr\rrVert _{2} \bigr],
\nonumber
\end{eqnarray}
where $(X,Y)$ and $(X',Y')$ are $\stackrel{\mathrm{i.i.d.}}{\sim}P_{XY}$.
As in the case of the energy distance, \citet{Lyons2011} established
that the generalization of the distance covariance is possible to
metric spaces of negative type. We extend this notion to semimetric
spaces of negative type.
%
\begin{defn}
Let $(\mathcal{X},\rho_{\mathcal{X}})$ and $(\mathcal{Y},\rho
_{\mathcal{Y}})$
be semimetric spaces of negative type, and let $X\sim P_{X}\in\mathcal
{M}_{\rho_{\mathcal{X}}}^{2}(\mathcal{X})$
and $Y\sim P_{Y}\in\mathcal{M}_{\rho_{\mathcal{Y}}}^{2}(\mathcal{Y})$,
having joint distribution $P_{XY}$. The generalized distance covariance
of $X$ and $Y$ is
%
%
\begin{eqnarray}\label{eqdCovintermsofdistances-1}
\mathcal{V}_{\rho_{\mathcal{X}},\rho_{\mathcal{Y}}}^{2}(X,Y) & = & \mathbb{E}_{XY}
\mathbb{E}_{X'Y'}\rho_{\mathcal{X}}\bigl(X,X'\bigr)\rho
_{\mathcal
{Y}}\bigl(Y,Y'\bigr)
\nonumber\\
& &{} + \mathbb{E}_{X}\mathbb{E}_{X'}\rho_{\mathcal{X}}
\bigl(X,X'\bigr)\mathbb {E}_{Y}\mathbb{E}_{Y'}
\rho_{\mathcal{Y}}\bigl(Y,Y'\bigr)
\\
& &{} - 2\mathbb{E}_{XY} \bigl[\mathbb{E}_{X'}
\rho_{\mathcal
{X}}\bigl(X,X'\bigr)\mathbb{E}_{Y'}
\rho_{\mathcal{Y}}\bigl(Y,Y'\bigr) \bigr].
\nonumber
\end{eqnarray}
\end{defn}
As with the energy distance, the moment conditions ensure that the
expectations are finite (which can be seen using the Cauchy--Schwarz
inequality). Equivalently, the generalized distance covariance can
be represented in integral form,
%
%
\begin{equation}
\mathcal{V}_{\rho_{\mathcal{X}},\rho_{\mathcal{Y}}}^{2}(X,Y)=\int \rho _{\mathcal{X}}
\rho_{\mathcal{Y}}\, d \bigl([P_{XY}-P_{X}P_{Y}]
\times [P_{XY}-P_{X}P_{Y}] \bigr),\label{eqdcovintegralform}
\end{equation}
where $\rho_{\mathcal{X}}\rho_{\mathcal{Y}}$ is viewed as a function
on $ (\mathcal{X}\times\mathcal{Y} )\times
(\mathcal
{X}\times\mathcal{Y} )$.
Furthermore, \citet{Lyons2011}, Theorem 3.20, shows that distance
covariance in a metric space characterizes independence\vspace*{1pt} [i.e.,
$\mathcal
{V}_{\rho_{\mathcal{X}},\rho_{\mathcal{Y}}}^{2}(X,Y)=0$
if and only if $X$ and $Y$ are independent] if the metrics $\rho
_{\mathcal{X}}$
and $\rho_{\mathcal{Y}}$ satisfy an additional property, termed \emph{strong
negative type}. The discussion of this property is relegated to Section~\ref{secDistinguishing-probability-distributions}.
%
\begin{rem}
\label{remdcov=00003Ddenergy?}While the form of \eqref{eqintegralexpressionenergydistance}
and \eqref{eqdcovintegralform} suggests that the energy distance
and the distance covariance are closely related, it is not clear whether
$\mathcal{V}_{\rho_{\mathcal{X}},\rho_{\mathcal{Y}}}^{2}(X,Y)$ is
simply $D_{E,\tilde{\rho}}(P_{XY},P_{X}P_{Y})$ for some semimetric
$\tilde{\rho}$ on $\mathcal{X}\times\mathcal{Y}$. In particular,
$-\rho_{\mathcal{X}}\rho_{\mathcal{Y}}$ is certainly not a semimetric.
This question will be resolved in Corollary \ref{cordcovandenergy}.
\end{rem}

\section{Kernel-based approach}\label{seckernel-based-approach}

In this section, we introduce concepts and notation required to understand
reproducing kernel Hilbert spaces (Section~\ref{subRKHS}), and distribution
embeddings into RKHS. We then introduce the maximum mean discrepancy
(MMD) and Hilbert--Schmidt independence criterion (HSIC).

\subsection{RKHS and kernel embeddings}
\label{subRKHS}

We begin with the definition of a reproducing kernel Hilbert space
(RKHS).
%
\begin{defn}[(RKHS)] Let $\mathcal{H}$ be a Hilbert space of real-valued
functions defined on $\mathcal{Z}$. A function $k\dvtx \mathcal{Z}\times
\mathcal{Z}\to\mathbb{R}$
is called \emph{a reproducing kernel} of $\mathcal{H}$ if:
\begin{longlist}[1.]
\item[1.]$\forall z\in\mathcal{Z}, k(\cdot,z)\in\mathcal{H}$, and
\item[2.]$\forall z\in\mathcal{Z}, \forall f\in\mathcal{H},
\langle
f,k(\cdot,z) \rangle_{\mathcal{H}}=f(z)$.
\end{longlist}

If $\mathcal{H}$ has a reproducing kernel, it is said to be \emph{a
reproducing kernel Hilbert space} (RKHS).
\end{defn}
According to the Moore--Aronszajn theorem [\citet{BerTho04}, page 19],
for every symmetric, positive definite function (henceforth \emph{kernel})
$k\dvtx \mathcal{Z}\times\mathcal{Z}\to\mathbb{R}$, there is an associated
RKHS $\mathcal{H}_{k}$ of real-valued functions on $\mathcal{Z}$
with reproducing kernel $k$. The map $\varphi\dvtx \mathcal{Z}\to\mathcal
{H}_{k}$,
$\varphi\dvtx z\mapsto k(\cdot,z)$ is called the canonical feature map
or the Aronszajn map of $k$. We will say that $k$ is a \emph{nondegenerate
kernel} if its Aronszajn map is injective. The notion of feature map
can be extended to kernel embeddings of finite signed Borel measures
on $\mathcal{Z}$ [\citet{SmoGreSonSch07,SriGreFukLanetal10}, \citet{BerTho04}, Chapter~4].
%
\begin{defn}[(Kernel embedding)] Let $k$ be a kernel on $\mathcal{Z}$,
and $\nu\in\mathcal{M}(\mathcal{Z})$. The \emph{kernel embedding}
of $\nu$ into the RKHS $\mathcal{H}_{k}$ is $\mu_{k}(\nu)\in
\mathcal{H}_{k}$
such that $\int f(z)\,d\nu(z)= \langle f,\mu_{k}(\nu)
\rangle
_{\mathcal{H}_{k}}$
for all $f\in\mathcal{H}_{k}$.
\end{defn}
Alternatively, the kernel embedding can be defined by the Bochner
integral $\mu_{k}(\nu)=\int k(\cdot,z) \,d\nu(z)$. If a measurable
kernel $k$ is a bounded function, $\mu_{k}(\nu)$ exists for all
$\nu\in\mathcal{M}(\mathcal{Z})$. On the other hand, if $k$ is not
bounded, there will always exist $\nu\in\mathcal{M}(\mathcal{Z})$,
for which $\int k(\cdot,z) \,d\nu(z)$ diverges. The kernels we will
consider in this paper will be continuous, and hence measurable, but
unbounded, so kernel embeddings will not be defined for some finite
signed measures. Thus, we need to restrict our attention to a particular
class of measures for which kernel embeddings exist (this will be
later shown to reflect the condition that random variables considered
in distance covariance tests must have finite moments). Let $k$ be
a measurable kernel on $\mathcal{Z}$, and denote, for $\theta>0$,
%
%
\begin{equation}
\mathcal{M}_{k}^{\theta}(\mathcal{Z})= \biggl\{ \nu\in\mathcal
{M}(\mathcal {Z}) \dvtx \int k^{\theta}(z,z) \,d\vert\nu\vert(z)<\infty \biggr
\}.
\end{equation}
Clearly,
%
%
\begin{equation}
\theta_{1}\leq\theta_{2}\quad\implies\quad\mathcal{M}_{k}^{\theta
_{2}}(
\mathcal {Z})\subseteq\mathcal{M}_{k}^{\theta_{1}}(\mathcal{Z}).
\end{equation}
Note that the kernel embedding $\mu_{k}(\nu)$ is well defined
$\forall
\nu\in\mathcal{M}_{k}^{1/2}(\mathcal{Z})$,
by the Riesz representation theorem.

\subsection{Maximum mean discrepancy}

As we have seen, kernel embeddings of Borel probability measures in
$\mathcal{M}_{+}^{1}(\mathcal{Z})\cap\mathcal{M}_{k}^{1/2}(\mathcal{Z})$
do exist, and we can introduce the notion of distance between Borel
probability measures in this set using the Hilbert space distance
between their embeddings.
%
\begin{defn}[(Maximum mean discrepancy)]
\label{defmmd} Let $k$ be a
kernel on $\mathcal{Z}$, and let $P,Q\in\mathcal{M}_{+}^{1}(\mathcal
{Z})\cap\mathcal{M}_{k}^{1/2}(\mathcal{Z})$.
The \emph{maximum mean discrepancy} (\textit{MMD}) $\gamma_{k}$ between $P$
and $Q$ is given by \citet{Gretton2012}, Lemma 4,
\[
\gamma_{k}(P,Q) =  \bigl\llVert \mu_{k}(P)-
\mu_{k}(Q)\bigr\rrVert _{\mathcal{H}_{k}}.
\]
\end{defn}
The following alternative representation of the squared MMD [from \citet{Gretton2012}, Lemma 6]
will be useful
%
%
\begin{eqnarray}\label{eqMMD}
\gamma_{k}^{2}(P,Q) & = & \mathbb{E}_{ZZ'}k
\bigl(Z,Z'\bigr)+\mathbb {E}_{WW'}k\bigl(W,W'
\bigr)-2\mathbb{E}_{ZW}k(Z,W)
\nonumber
\\[-8pt]
\\[-8pt]
\nonumber
& = & \int\int k\, d \bigl( [P-Q ]\times [P-Q ] \bigr),
\end{eqnarray}
where $Z,Z'\stackrel{\mathrm{i.i.d.}}{\sim}P$ and $W,W'\stackrel{\mathrm{i.i.d.}}{\sim}Q$.
If the restriction of $\mu_{k}$ to some $\mathcal{P}(\mathcal
{Z})\subseteq\mathcal{M}_{+}^{1}(\mathcal{Z})$
is well defined and injective, then $k$ is said to be characteristic
to $\mathcal{P}(\mathcal{Z})$, and it is said to be characteristic
(without further qualification) if it is characteristic to $\mathcal
{M}_{+}^{1}(\mathcal{Z})$.
When $k$ is characteristic, $\gamma_{k}$ is a metric on the entire
$\mathcal{M}_{+}^{1}(\mathcal{Z})$, that is, $\gamma_{k}
(P,Q )=0$
iff $P=Q$, $\forall P,Q\in\mathcal{M}_{+}^{1}(\mathcal{Z})$. Conditions
under which kernels are characteristic have been studied by \citet
{SriGreFukLanetal08,FukSriGreSch09,SriGreFukLanetal10}.
An alternative interpretation of \eqref{eqMMD} is as an integral
probability metric [\citet{Mueller97}],
%
%
\begin{equation}
\gamma_{k}(P,Q)=\sup_{f\in\mathcal{H}_{k},\llVert  f\rrVert
_{\mathcal{H}_{k}}\leq1} \bigl[
\mathbb{E}_{Z\sim P}f(Z)-\mathbb {E}_{W\sim Q}f(W) \bigr].
\end{equation}
See \citet{Gretton2012} and \citet{SriFukGreSchetal11} for
details.

\subsection{Hilbert--Schmidt independence criterion (HSIC)}

The MMD can be employed to measure statistical dependence between
random variables [Gretton et~al. (\citeyear{GreBouSmoSch05,GreFukTeoSonetal08}), \citet{SmoGreSonSch07,GreGyo10,Zhang2011}].
Let $\mathcal{X}$ and $\mathcal{Y}$ be two nonempty topological
spaces and let $k_{\mathcal{X}}$ and $k_{\mathcal{Y}}$ be kernels
on $\mathcal{X}$ and $\mathcal{Y}$, with respective RKHSs $\mathcal
{H}_{k_{\mathcal{X}}}$
and $\mathcal{H}_{k_{\mathcal{Y}}}$. Then, by applying \citeauthor{Steinwart2008book} [(\citeyear{Steinwart2008book}), Lemma
4.6, page~114],
%
%
\begin{equation}
k \bigl( (x,y ), \bigl(x',y' \bigr)
\bigr)=k_{\mathcal
{X}}\bigl(x,x'\bigr)k_{\mathcal{Y}}
\bigl(y,y'\bigr)\label{eqproductkernel}
\end{equation}
is a kernel on the product space $\mathcal{X}\times\mathcal{Y}$
with RKHS $\mathcal{H}_{k}$ isometrically isomorphic to the tensor
product $\mathcal{H}_{k_{\mathcal{X}}}\otimes\mathcal
{H}_{k_{\mathcal{Y}}}$.
%
\begin{defn}
Let $X\sim P_{X}$ and $Y\sim P_{Y}$ be random variables on $\mathcal{X}$
and~$\mathcal{Y}$, respectively, having joint distribution $P_{XY}$.
Furthermore, let $k$ be a kernel on $\mathcal{X}\times\mathcal{Y}$,
given in \eqref{eqproductkernel}. The Hilbert--Schmidt independence
criterion (HSIC) of $X$ and $Y$ is the MMD $\gamma_{k}$ between
the joint distribution $P_{XY}$ and the product of its marginals
$P_{X}P_{Y}$.
\end{defn}
Following \citet{SmoGreSonSch07}, Section~2.3, we can expand HSIC
as
%
%
\begin{eqnarray}\label{eqHSICexpansion}
& & \gamma_{k}^{2}(P_{XY},P_{X}P_{Y})
\nonumber
\\
&&\qquad =  \bigl\llVert \mathbb{E}_{XY} \bigl[k_{\mathcal{X}}(\cdot,X)
\otimes k_{\mathcal{Y}}(\cdot,Y) \bigr]-\mathbb{E}_{X}k_{\mathcal
{X}}(
\cdot,X)\otimes\mathbb{E}_{Y}k_{\mathcal{Y}}(\cdot,Y)\bigr\rrVert
_{\mathcal
{H}_{k_{\mathcal{X}}}\otimes\mathcal{H}_{k_{\mathcal
{Y}}}}^{2}
\nonumber
\\[-8pt]
\\[-8pt]
\nonumber
&&\qquad =  \mathbb{E}_{XY}\mathbb{E}_{X'Y'}k_{\mathcal
{X}}
\bigl(X,X'\bigr)k_{\mathcal
{Y}}\bigl(Y,Y'\bigr)+
\mathbb{E}_{X}\mathbb{E}_{X'}k_{\mathcal{X}}
\bigl(X,X'\bigr)\mathbb {E}_{Y}\mathbb{E}_{Y'}k_{\mathcal{Y}}
\bigl(Y,Y'\bigr)
\\
& &\qquad\quad{} - 2\mathbb{E}_{X'Y'} \bigl[\mathbb{E}_{X}k_{\mathcal
{X}}
\bigl(X,X'\bigr)\mathbb{E}_{Y}k_{\mathcal{Y}}
\bigl(Y,Y'\bigr) \bigr].
\nonumber
\end{eqnarray}
It can be shown that this quantity is equal to the
squared Hilbert--Schmidt norm of the covariance operator between RKHSs
[\citet{GreBouSmoSch05}]. We claim that $\gamma_{k}^{2}(P_{XY},P_{X}P_{Y})$
is well defined as long as $P_{X}\in\mathcal{M}_{k_{\mathcal
{X}}}^{1}(\mathcal{X})$
and $P_{Y}\in\mathcal{M}_{k_{\mathcal{Y}}}^{1}(\mathcal{Y})$. Indeed,
this is a sufficient condition for $\mu_{k}(P_{XY})$ to exist, since
it implies that $P_{XY}\in\mathcal{M}_{k}^{1/2}(\mathcal{X}\times
\mathcal{Y})$,
which can be seen from the Cauchy--Schwarz inequality,
\begin{eqnarray*}
&&\int k^{1/2} \bigl( (x,y ), (x,y ) \bigr)\,dP_{XY}(x,y)\\[-2pt]
&&\qquad=\int
k_{\mathcal{X}}^{1/2}(x,x)k_{\mathcal{Y}}^{1/2}(y,y)
\,dP_{XY}(x,y)
\\[-2pt]
&&\qquad\leq \biggl(\int k_{\mathcal{X}}(x,x)\,dP_{X}(x)\int
k_{\mathcal
{Y}}(y,y)\,dP_{Y}(y) \biggr)^{1/2}.
\end{eqnarray*}
Furthermore, the embedding $\mu_{k}(P_{X}P_{Y})$ of the product of
marginals also exists, as it can be identified with the tensor product
$\mu_{k_{\mathcal{X}}}(P_{X})\otimes\mu_{k_{\mathcal{Y}}}(P_{Y})$,
where $\mu_{k_{\mathcal{X}}}(P_{X})$ exists since $P_{X}\in\mathcal
{M}_{k_{\mathcal{X}}}^{1}(\mathcal{X})\subset\mathcal
{M}_{k_{\mathcal
{X}}}^{1/2}(\mathcal{X})$,
and $\mu_{k_{\mathcal{Y}}}(P_{Y})$ exists since $P_{Y}\in\mathcal
{M}_{k_{\mathcal{Y}}}^{1}(\mathcal{Y})\subset\mathcal
{M}_{k_{\mathcal
{Y}}}^{1/2}(\mathcal{Y})$.

\section{Correspondence between kernels
and semimetrics}\label{secCorrespondencePDKvsNTsM}

In this section, we develop the correspondence of semimetrics of negative
type (Section~\ref{subSemimetrics-negative-type}) to the RKHS theory,
that is, to symmetric positive definite kernels. This correspondence
will be key to proving the equivalence between the energy distance
and MMD, and the equivalence between distance covariance and HSIC
in Section~\ref{secEquivalences}.

\subsection{Distance-induced kernels}
\label{subDistance-kernels}

Semimetrics of negative type and symmetric positive definite kernels
are closely related, as summarized in the following lemma, adapted
from \citet{BerChrRes84}, Lemma~2.1, page~74.
%
\begin{lem}
\label{lemkernel-from-semimetric}Let $\mathcal{Z}$ be a nonempty
set, and $\rho\dvtx \mathcal{Z}\times\mathcal{Z}\to\mathbb{R}$ a semimetric
on $\mathcal{Z}$. Let $z_{0}\in\mathcal{Z}$, and denote
$k(z,z')=\rho
(z,z_{0})+\rho(z',z_{0})-\rho(z,z')$.
Then $k$ is positive definite if and only if $\rho$ satisfies \eqref
{eqCND}.
\end{lem}
As a consequence, $k(z,z')$ defined above is a valid kernel on
$\mathcal{Z}$
whenever $\rho$ is a semimetric of negative type. For convenience,
we will work with such kernels scaled by $1/2$.
%
\begin{defn}[(Distance-induced kernel)] Let $\rho$ be a semimetric of
negative type on $\mathcal{Z}$ and let $z_{0}\in\mathcal{Z}$. The
kernel
%
%
\begin{equation}
k\bigl(z,z'\bigr)=\tfrac{1}{2} \bigl[\rho(z,z_{0})+
\rho\bigl(z',z_{0}\bigr)-\rho \bigl(z,z'
\bigr) \bigr]\label{eqdistancekernel}
\end{equation}
is said to be the \emph{distance-induced kernel} induced by $\rho$
and \emph{centred} at $z_{0}$.
\end{defn}\vspace*{-12pt}\eject
For brevity, we will drop ``induced'' hereafter, and say that $k$
is simply the \emph{distance kernel} (with some abuse of terminology).
Note that distance kernels are not strictly positive definite, that is,
it is not true that $\forall n\in\mathbb{N}$, and for distinct
$z_{1},\ldots,z_{n}\in\mathcal{Z}$,
\[
\sum_{i=1}^{n}\sum
_{j=1}^{n}\alpha_{i}\alpha
_{j}k(z_{i},z_{j})=0\quad\implies\quad
\alpha_{i}=0\ \forall i.
\]
Indeed, if $k$ were given by \eqref{eqdistancekernel}, it would
suffice to take $n=1$, since $k(z_{0},z_{0})=0$. By varying the
point at the center $z_{0}$, we obtain a family
\[
\mathcal{K}_{\rho}= \bigl\{ \tfrac{1}{2} \bigl[
\rho(z,z_{0})+\rho \bigl(z',z_{0}\bigr)-\rho
\bigl(z,z'\bigr) \bigr] \bigr\} _{z_{0}\in\mathcal{Z}}
\]
of distance kernels induced by $\rho$. The following proposition
follows readily from the definition of $\mathcal{K}_{\rho}$ and shows
that one can always express \eqref{eqrhoviaHilbert} from Proposition~\ref{prosemimetrichilbertianmetric} in terms of the canonical
feature map for the RKHS $\mathcal{H}_{k}$.
%
\begin{prop}
\label{propropertiesofKrho}Let $(\mathcal{Z},\rho)$ be a semimetric
space of negative type, and $k\in\mathcal{K}_{\rho}$. Then:
\begin{longlist}[1.]
\item[1.]$\rho(z,z')=k(z,z)+k(z',z')-2k(z,z')=\llVert
k(\cdot
,z)-k(\cdot,z')\rrVert _{\mathcal{H}_{k}}^{2}$.
\item[2.]$k$ is nondegenerate, that is, the Aronszajn map $z\mapsto
k(\cdot,z)$
is injective.
\end{longlist}
\end{prop}

\begin{example}
\label{exavariousexponents}Let $\mathcal{Z}\subseteq\mathbb{R}^{d}$
and write $\rho_{q}(z,z')=\llVert  z-z'\rrVert ^{q}$. By
Proposition~\ref{prosemimetrichilbertianmetric}, $\rho_{q}$ is a valid semimetric
of negative type for $0<q\leq2$. The corresponding kernel centered
at $z_{0}=0$ is given by the covariance function of the fractional
Brownian motion,
%
%
\begin{equation}
k_{q}\bigl(z,z'\bigr)=\tfrac{1}{2} \bigl(\llVert z
\rrVert ^{q}+\bigl\llVert z'\bigr\rrVert ^{q}-
\bigl\llVert z-z'\bigr\rrVert ^{q} \bigr).\label{eqeuclideankernel}
\end{equation}
\end{example}
Note that while \citeauthor{Lyons2011} [(\citeyear{Lyons2011}), page 9] also uses the results in
Proposition \ref{prosemimetrichilbertianmetric} to characterize
metrics of negative type using embeddings to general Hilbert spaces,
the relation with the theory of reproducing kernel Hilbert spaces
is not exploited in his work.

\subsection{Semimetrics generated by kernels}

We now further develop the link between semimetrics of negative type
and kernels. We start with a simple corollary of Proposition \ref
{prosemimetrichilbertianmetric}.
%
\begin{cor}
Let $k$ be any nondegenerate kernel on $\mathcal{Z}$. Then,
%
%
\begin{equation}
\rho\bigl(z,z'\bigr)=k(z,z)+k\bigl(z',z'
\bigr)-2k\bigl(z,z'\bigr)\label{eqrhoFromK}
\end{equation}
defines a valid semimetric $\rho$ of negative type on $\mathcal{Z}$.
\end{cor}
%
\begin{defn}[(Equivalent kernels)]
Whenever the kernel $k$ and semimetric~$\rho$ satisfy \eqref{eqrhoFromK}, we will say that $k$ \emph{generates}
$\rho$. If two kernels generate the same semimetric, we will say
that they are \emph{equivalent kernels}.
\end{defn}
It is clear that every distance kernel $\tilde{k}\in\mathcal
{K}_{\rho}$
induced by $\rho$, also generates $\rho$. However, there are many
other kernels that generate $\rho$. The following
proposition is straightforward to show and gives a condition under
which two kernels are equivalent.
%
\begin{prop}
\label{propkernelequivalence}Let $k$ and $\tilde{k}$ be two kernels
on $\mathcal{Z}$. $k$ and $\tilde{k}$ are equivalent if and only
if $\tilde{k}(z,z')=k(z,z')+f(z)+f(z')$, for some shift function
$f\dvtx \mathcal{Z}\to\mathbb{R}$.
\end{prop}
Not every choice of shift function $f$ in Proposition
\ref{propkernelequivalence} will be valid, as both $k$ and $\tilde{k}$
are required to be positive definite. An important class of shift
functions can be derived using RKHS functions, however. Namely, let
$k$ be a kernel on $\mathcal{Z}$ and let $f\in\mathcal{H}_{k}$,
and define a kernel
\begin{eqnarray*}
\tilde{k}_{f}\bigl(z,z'\bigr) & = & \bigl\langle k(
\cdot,z)-f,k\bigl(\cdot,z'\bigr)-f \bigr\rangle_{\mathcal{H}_{k}}
\\
& = & k\bigl(z,z'\bigr)-f(z)-f\bigl(z'\bigr)+\llVert f
\rrVert _{\mathcal{H}_{k}}^{2}.
\end{eqnarray*}
Since it is representable as an inner product in a Hilbert space,
$\tilde{k}_{f}$ is a valid kernel which is equivalent to $k$ by
Proposition \ref{propkernelequivalence}. As a special case, if
$f=\mu_{k}(P)$ for some $P\in\mathcal{M}_{+}^{1}(\mathcal{Z})$,
we obtain the kernel centred at probability measure $P$:
%
%
\begin{equation}\qquad
\tilde{k}_{P}\bigl(z,z'\bigr):=k\bigl(z,z'
\bigr)+\mathbb{E}_{WW'}k\bigl(W,W'\bigr)-\mathbb
{E}_{W}k(z,W)-\mathbb{E}_{W}k\bigl(z',W\bigr),
\label{eqcentredkernels}
\end{equation}
with $W,W'\stackrel{\mathrm{i.i.d.}}{\sim}P$. Note that $\mathbb
{E}_{ZZ'\stackrel
{\mathrm{i.i.d.}}{\sim}P}\tilde{k}_{P}(Z,Z')=0$,
that is, $\mu_{\tilde{k}_{P}}(P)=0$. The kernels of form \eqref{eqcentredkernels}
that are centred at the point masses $P=\delta_{z_{0}}$ are precisely
the distance kernels equivalent to $k$.

\begin{figure}

\includegraphics{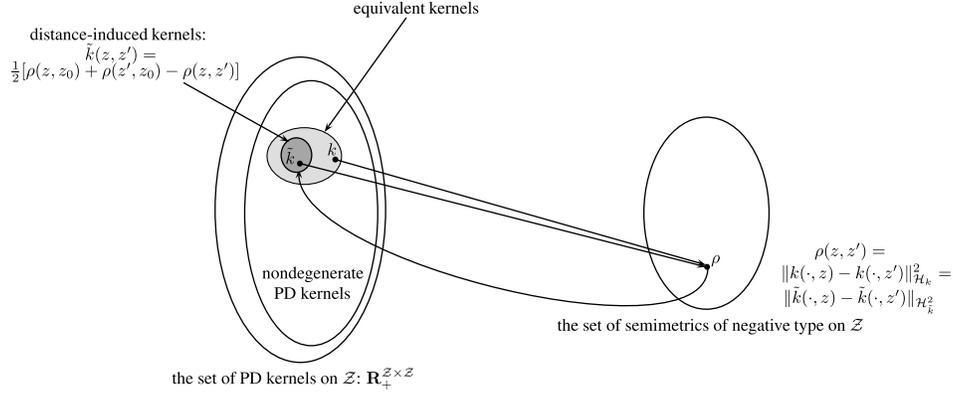}

\caption{The relationship between kernels and semimetrics.
An equivalence class of nondegenerate PD kernels is associated to
a single semimetric of negative type, and distance kernels induced
by that semimetric form only a subset of that class.}\label{figdiagram}
\end{figure}

The relationship between positive definite kernels and semimetrics
of negative type is illustrated in Figure~\ref{figdiagram}.
%
\begin{rem}
The requirement that kernels be characteristic (as introduced below
Definition \ref{defmmd}) is clearly important in hypothesis testing.
A second family of kernels, widely used in the machine learning literature,
are the universal kernels: universality can be used to guarantee consistency
of learning algorithms [\citet{Steinwart2008book}]. While these two
notions are closely related, and in some cases coincide [\citet
{Sriperumbudur2011}],
one can easily construct nonuniversal characteristic kernels as a
consequence of Proposition \ref{propkernelequivalence}. See
{Appendix}~\ref{secLink-universal-kernels}
for details.
\end{rem}

\subsection{Existence of kernel embedding through a semimetric}
In Section~\ref{subRKHS}, we have seen that a sufficient condition
for the kernel embedding $\mu_{k}(\nu)$ of $\nu\in\mathcal
{M}(\mathcal{Z})$
to exist is that $\nu\in\mathcal{M}_{k}^{1/2}(\mathcal{Z})$. We will
now interpret this condition in terms of the semimetric $\rho$ generated
by $k$, by relating $\mathcal{M}_{k}^{\theta}(\mathcal{Z})$ to the
space $\mathcal{M}_{\rho}^{\theta}(\mathcal{Z})$ of measures with
finite $\theta$-moment w.r.t. $\rho$.
%
\begin{prop}
\label{propequalityofspacesofmeasures}Let $k$ be a kernel that
generates semimetric $\rho$, and let $n\in\mathbb{N}$. Then
$\mathcal
{M}_{k}^{n/2}(\mathcal{Z})=\mathcal{M}_{\rho}^{n/2}(\mathcal{Z})$.
In particular, if $k_{1}$ and $k_{2}$ generate the same semimetric
$\rho$, then $\mathcal{M}_{k_{1}}^{n/2}(\mathcal
{Z})=\mathcal
{M}_{k_{2}}^{n/2}(\mathcal{Z})$.
\end{prop}
\begin{pf}
Let $\theta\ge\frac{1}{2}$. Suppose $\nu\in\mathcal
{M}_{k}^{\theta
}(\mathcal{Z})$.
Then we have
\begin{eqnarray*}
\int\rho^{\theta}(z,z_{0}) \,d|\nu|(z) &  =&\int\bigl\Vert k(\cdot
,z)-k(\cdot,z_{0})\bigr\Vert_{\mathcal{H}_{k}}^{2\theta} \,d|\nu|(z)
\\
& \le&\int \bigl(\bigl\Vert k(\cdot,z)\bigr\Vert_{\mathcal{H}_{k}}+\bigl\Vert k(
\cdot,z_{0})\bigr\Vert_{\mathcal{H}_{k}} \bigr)^{2\theta} \,d|\nu|(z)
\\
&  \le&2^{2\theta-1} \biggl(\int\bigl\Vert k(\cdot,z)\bigr\Vert_{\mathcal
{H}_{k}}^{2\theta}
\,d|\nu|(z)+\int\bigl\Vert k(\cdot,z_{0})\bigr\Vert _{\mathcal
{H}_{k}}^{2\theta}
\,d|\nu|(z) \biggr)
\\
&  =&2^{2\theta-1} \biggl(\int k^{\theta}(z,z) \,d|
\nu|(z)+k^{\theta
}(z_{0},z_{0})|\nu|(\Cal{Z}) \biggr)
\\
&  <& \infty,
\end{eqnarray*}
where we have used that $a^{2\theta}$ is a convex function of $a$.
From the above it is clear that $\Cal{M}_{k}^{\theta}(\Cal
{Z})\subset
\Cal{M}_{\rho}^{\theta}(\Cal{Z})$
for $\theta\ge1/2$.

To prove the other direction, we show by induction that $\Cal{M}_{\rho
}^{\theta}(\Cal{Z})\subset\Cal{M}_{k}^{n/2}(\Cal{Z})$
for $\theta\ge\frac{n}{2}$, $n\in\bb{N}$. Let $n=1$, $\theta\ge
\frac{1}{2}$,
and suppose that $\nu\in\Cal{M}_{\rho}^{\theta}(\Cal{X})$. Then,
by invoking the reverse triangle and Jensen's inequalities, we have
\begin{eqnarray*}
\int\rho^{\theta}(z,z_{0})\,d\llvert \nu\rrvert (z)&=&\int
\bigl\llVert k(\cdot,z)-k(\cdot,z_{0})\bigr\rrVert _{\mathcal{H}_{k}}^{2\theta
}\,d|
\nu |(z)
\\
& \geq&\int\bigl\llvert k^{1/2}(z,z)-k^{1/2}(z_{0},z_{0})
\bigr\rrvert ^{2\theta
}\,d|\nu|(z)
\\
& \geq& \biggl|\int k^{1/2}(z,z) \,d|\nu|(z)-\llVert \nu\rrVert
_{\mathrm{TV}}k^{1/2}(z_{0},z_{0})
\biggr|^{2\theta},
\end{eqnarray*}
which implies $\nu\in\Cal{M}_{k}^{1/2}(\Cal{Z})$, thereby satisfying
the result for $n=1$. Suppose the result holds for $\theta\ge\frac{n-1}{2}$,
that is, $\Cal{M}_{\rho}^{\theta}(\Cal{Z})\subset\Cal
{M}_{k}^{(n-1)/2}(\Cal{Z})$
for $\theta\ge\frac{n-1}{2}$. Let $\nu\in\Cal{M}_{\rho}^{\theta
}(\Cal{Z})$
for $\theta\ge\frac{n}{2}$. Then we have
\begin{eqnarray*}
&& \int\rho^{\theta}(z,z_{0}) \,d|\nu|(z)\\
&&\qquad= \int \bigl(\bigl\Vert k(\cdot
,z)-k(\cdot,z_{0})\bigr\Vert_{\mathcal{H}_{k}}^{n}
\bigr)^{{2\theta
}/{n}} \,d|\nu|(z)
\\
&&\qquad \ge\biggl\llvert \int \bigl(\bigl\Vert k(\cdot,z)\bigr\Vert_{\mathcal
{H}_{k}}-\bigl\Vert k(
\cdot,z_{0})\bigr\Vert_{\mathcal{H}_{k}} \bigr)^{n} \,d|\nu|(z)\biggr
\rrvert ^{
{2\theta}/{n}}
\\
&&\qquad = \Biggl\llvert \int\sum_{r=0}^{n}(-1)^{r}
\pmatrix{ n
\cr
r }\bigl\Vert k(\cdot,z)\bigr\Vert_{\mathcal{H}_{k}}^{n-r}
\bigl\Vert k(\cdot,z_{0})\bigr\Vert_{\mathcal{H}_{k}}^{r} \,d|\nu|(z)
\Biggr\rrvert ^{{2\theta
}/{n}}
\\
&&\qquad = \Biggl|\underbrace{\int k^{{n}/{2}}(z,z) \,d|\nu|(z)}_{A}\\
&&\hspace*{2pt}\qquad\quad{}+
\underbrace{\sum_{r=1}^{n}(-1)^{r}
\pmatrix{ n
\cr
r }k^{{r}/{2}}(z_{0},z_{0})
\int k^{{(n-r)}/{2}}(z,z) \,d|\nu |(z)}_{B} \Biggr|^{{2\theta}/{n}}.
\end{eqnarray*}
Note that the terms in $B$ are finite since for $\theta\ge\frac
{n}{2}\ge
\frac{n-1}{2}\ge\cdots\ge\frac{1}{2}$,
we have $\Cal{M}_{\rho}^{\theta}(\Cal{Z})\subset\Cal
{M}_{k}^{(n-1)/2}(\Cal{Z})\subset\cdots\subset\Cal{M}_{k}^{1}(\Cal
{Z})\subset\Cal{M}_{k}^{1/2}(\Cal{Z})$
and therefore $A$ is finite, which means $\nu\in\Cal
{M}_{k}^{n/2}(\Cal{Z})$,
that is, $\Cal{M}_{\rho}^{\theta}(\Cal{Z})\subset\Cal
{M}_{k}^{n/2}(\Cal{Z})$
for $\theta\ge\frac{n}{2}$. The result shows that $\Cal{M}_{\rho
}^{\theta}(\Cal{Z})=\Cal{M}_{k}^{\theta}(\Cal{Z})$
for all $\theta\in\{\frac{n}{2}\dvtx n\in\bb{N}\}$.
\end{pf}
%
\begin{rem}
\label{remark-sufficiency-of-moment-condition}We are now able to
show that $P,Q\in\mathcal{M}_{\rho}^{1}(\mathcal{Z})$ is sufficient
for the existence of $D_{E,\rho}(P,Q)$, that is, to show validity of
Definition \ref{defenergydistance} for general semimetrics of
negative type $\rho$. Namely, we let $k$ be any kernel that generates
$\rho$, whereby $P,Q\in\mathcal{M}_{k}^{1}(\mathcal{Z})$. Thus,
\[
\mathbb{E}_{ZW}\rho(Z,W) =  \mathbb{E}_{Z}k(Z,Z)+\mathbb
{E}_{W}k(W,W)-2\mathbb{E}_{ZW}k(Z,W)<\infty,
\]
where the first term is finite as $P\in\mathcal{M}_{k}^{1}(\mathcal{Z})$,
the second term is finite as $Q\in\mathcal{M}_{k}^{1}(\mathcal{Z})$,
and the third term is finite by noticing that $\llvert k(z,w)\rrvert \leq
k^{1/2}(z,z)\times k^{1/2}(w,w)$
and $P,Q\in\mathcal{M}_{k}^{1}(\mathcal{Z})\subset\mathcal
{M}_{k}^{1/2}(\mathcal{Z})$.
\end{rem}
Proposition \ref{propequalityofspacesofmeasures} gives a natural
interpretation of conditions on probability measures in terms of moments
w.r.t. $\rho$. Namely, the kernel embedding $\mu_{k}(P)$, where
kernel $k$ generates the semimetric $\rho$, exists for every $P$
with finite half-moment w.r.t.~$\rho$, and thus the MMD, $\gamma_{k}(P,Q)$
between $P$ and $Q$ is well defined whenever both $P$ and $Q$
have finite half-moments w.r.t.~$\rho$. Furthermore, HSIC between
random variables $X$ and $Y$ is well defined whenever their marginals
$P_{X}$ and $P_{Y}$ have finite first moments w.r.t. semimetric
$\rho_{\mathcal{X}}$ and $\rho_{\mathcal{Y}}$ generated by kernels
$k_{\mathcal{X}}$ and $k_{\mathcal{Y}}$ on their respective domains
$\mathcal{X}$ and $\mathcal{Y}$.

\section{Main results}\label{secEquivalences}

In this section, we establish the equivalence between the distance-based
approach and the RKHS-based approach to two-sample and independence
testing from Sections~\ref{secDistance-based-approach} and \ref{seckernel-based-approach},
respectively.

\subsection{Equivalence of MMD
and energy
distance}
\label{subenergydistancewithkernels}

\begin{figure}

\includegraphics{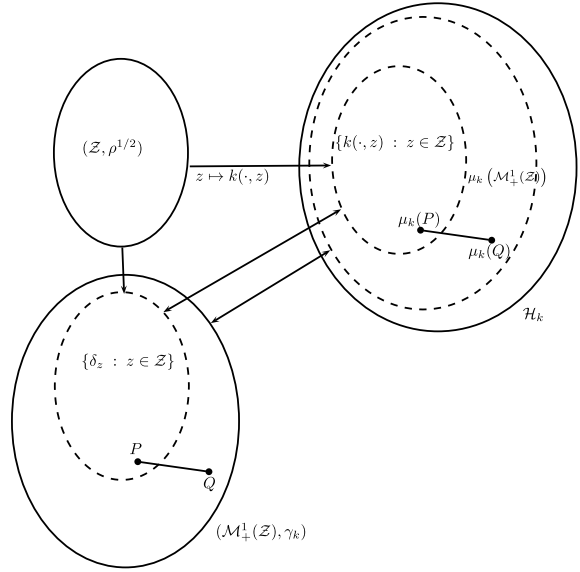}

\caption{Isometries relating the semimetric $\rho$ on
$\mathcal{Z}$ with the RKHS corresponding to a kernel $k$ that generates
$\rho$, and with the set of probability measures on $\mathcal{Z}$:
(1) $z\mapsto k(\cdot,z)$ embeds $ (\mathcal{Z},\rho
^{1/2} )$
into $\mathcal{H}_{k}$, (2) $z\mapsto\delta_{z}$ embeds $
(\mathcal
{Z},\rho^{1/2} )$
into $(\mathcal{M}_{+}^{1}(\mathcal{Z}),\gamma_{k})$, and (3)
$P\mapsto
\mu_{k}(P)$
embeds $(\mathcal{M}_{+}^{1}(\mathcal{Z}),\gamma_{k})$ into
$\mathcal
{H}_{k}$.}\label{figisometry}
\end{figure}

We show that for every $\rho$, the energy distance $D_{E,\rho}$
is related to the MMD associated to a kernel $k$ that generates~$\rho$.
%
\begin{thmm}
\label{thmm2sampledkern}Let $(\mathcal{Z},\rho)$ be a semimetric
space of negative type and let $k$ be any kernel that generates $\rho$.
Then
\[
D_{E,\rho}(P,Q)=2\gamma_{k}^{2}(P,Q)\qquad \forall P,Q\in
\mathcal {M}_{+}^{1}(\mathcal{Z})\cap\mathcal{M}_{\rho}^{1}(
\mathcal{Z}).
\]
In particular, equivalent kernels have the same maximum mean discrepancy.
\end{thmm}
\begin{pf}
Since $k$ generates $\rho$, we can write $\rho(z,w)=k(z,z)+k(w,w)-2k(z,w)$.
Denote $\nu=P-Q$. Then
\begin{eqnarray*}
D_{E,\rho}(P,Q) & = & -\int\int \bigl[k(z,z)+k(w,w)-2k(z,w) \bigr] \,d\nu
(z) \,d\nu(w)
\\
& = & 2\int\int k(z,w) \,d\nu(z) \,d\nu(w)
\\
& = & 2\gamma_{k}^{2}(P,Q),
\end{eqnarray*}
where we used the fact that $\nu(\mathcal{Z})=0$.
\end{pf}
This result may be compared with that of \citeauthor{Lyons2011} [(\citeyear{Lyons2011}), page 11,
equation~(3.9)]
for embeddings into general Hilbert spaces, where we have provided
the link to RKHS-based statistics (and MMD in particular).
Theorem~\ref{thmm2sampledkern} shows that all kernels that generate the
same semimetric $\rho$ on $\mathcal{Z}$ give rise to the same metric
$\gamma_{k}$ on (possibly a subset of) $\mathcal{M}_{+}^{1}(\mathcal{Z})$,
whence $\gamma_{k}$ is merely an extension of the metric induced
by $\rho^{1/2}$ on point masses, since
\[
\gamma_{k}(\delta_{z},\delta_{z'})=\bigl
\llVert k(\cdot,z)-k\bigl(\cdot,z'\bigr)\bigr\rrVert
_{\mathcal{H}_{k}}=\rho^{1/2}\bigl(z,z'\bigr).
\]
In other words, whenever kernel $k$ generates $\rho$, $z\mapsto\delta_{z}$
is an isometry between $(\mathcal{Z},\rho^{1/2})$ and $ \{
\delta
_{z}\dvtx z\in\mathcal{Z} \} \subset\mathcal{M}_{+}^{1}(\mathcal{Z})$,
endowed with the MMD metric $\gamma_{k}=\frac{1}{2}D_{E,\rho}^{1/2}$;
and the Aronszajn map $z\mapsto k(\cdot,z)$ is an isometric embedding
of a metric space $(\mathcal{Z},\rho^{1/2})$ into $\mathcal{H}_{k}$.
These isometries are depicted in Figure~\ref{figisometry}. For simplicity,
we show the case of a bounded kernel, where kernel embeddings are
well defined for all $P\in\mathcal{M}_{+}^{1}(\mathcal{Z})$, in which
case $(\mathcal{M}_{+}^{1}(\mathcal{Z}),\gamma_{k})$ and $\mu
_{k}
(\mathcal{M}_{+}^{1}(\mathcal{Z}) )=\{\mu_{k}(P) \dvtx P\in
\mathcal
{M}_{+}^{1}(\mathcal{Z})\}$
endowed with the Hilbert-space metric inherited from $\mathcal{H}_{k}$
are also isometric (note that this implies that the subsets of RKHSs
corresponding to equivalent kernels are also isometric).

\begin{rem}
\label{remfinitenessofmoments}Theorem \ref{thmm2sampledkern}
requires that $P,Q\in\mathcal{M}_{\rho}^{1}(\mathcal{Z})$, that is,
that $P$ and $Q$ have finite first moments w.r.t. $\rho$, as otherwise
the energy distance between $P$ and $Q$ may be undefined; for example,
each of the expectations $\mathbb{E}_{ZZ'}\rho(Z,Z')$, $\mathbb
{E}_{WW'}\rho(W,W')$
and $\mathbb{E}_{ZW}\rho(Z,W)$ may be infinite. However, as long
as a \emph{weaker condition} $P,Q\in\mathcal{M}_{\rho
}^{1/2}(\mathcal{Z})$
is satisfied, that is, $P$ and $Q$ have finite \emph{half}-moments
w.r.t. $\rho$, the maximum mean discrepancy $\gamma_{k}$ will be
well defined. If, in addition, $P,Q\in\mathcal{M}_{\rho
}^{1}(\mathcal{Z})$,
then the energy distance between $P$ and $Q$ is also well defined,
and must be equal to $\gamma_{k}$. We will later invoke the same
condition $P,Q\in\mathcal{M}_{k}^{1}(\mathcal{Z})$ when describing
the asymptotic distribution of the empirical maximum mean discrepancy
in Section~\ref{secConsistency}.
\end{rem}

\subsection{Equivalence between HSIC and distance
covariance}\label{subdcovwithkernels}

We now show that distance covariance is an instance of the Hilbert--Schmidt
independence criterion.
%
\begin{thmm}
\label{thmmdcovkern}Let $(\mathcal{X},\rho_{\mathcal{X}})$ and
$(\mathcal{Y},\rho_{\mathcal{Y}})$ be semimetric spaces of negative
type, and let $X\sim P_{X}\in\mathcal{M}_{\rho_{\mathcal
{X}}}^{2}(\mathcal{X})$
and $Y\sim P_{Y}\in\mathcal{M}_{\rho_{\mathcal{Y}}}^{2}(\mathcal{Y})$,
having joint distribution $P_{XY}$. Let $k_{\mathcal{X}}$ and
$k_{\mathcal{Y}}$
be any two kernels on $\mathcal{X}$ and $\mathcal{Y}$ that generate
$\rho_{\mathcal{X}}$ and $\rho_{\mathcal{Y}}$, respectively, and
denote
%
%
\begin{equation}
k \bigl( (x,y ), \bigl(x',y' \bigr)
\bigr)=k_{\mathcal
{X}}\bigl(x,x'\bigr)k_{\mathcal{Y}}
\bigl(y,y'\bigr).\label{eqtensorproductk}
\end{equation}
Then, $\mathcal{V}_{\rho_{\mathcal{X},}\rho_{\mathcal
{Y}}}^{2}(X,Y)=4\gamma_{k}^{2}(P_{XY},P_{X}P_{Y})$.
\end{thmm}
\begin{pf}
Define $\nu=P_{XY}-P_{X}P_{Y}$. Then
\begin{eqnarray*}
\mathcal{V}_{\rho_{\mathcal{X}},\rho_{\mathcal{Y}}}^{2}(X,Y) & = & \int \int\rho_{\mathcal{X}}
\bigl(x,x'\bigr)\rho_{\mathcal{Y}}\bigl(y,y'\bigr) \,d
\nu(x,y) \,d\nu \bigl(x',y'\bigr)
\\
& = & \int\int \bigl(k_{\mathcal{X}}(x,x)+k_{\mathcal
{X}}\bigl(x',x'
\bigr)-2k_{\mathcal{X}}\bigl(x,x'\bigr) \bigr)
\\
& &\hphantom{\int\int}\times{} \bigl(k_{\mathcal{Y}}(y,y)+k_{\mathcal{Y}}\bigl(y',y'
\bigr)-2k_{\mathcal
{Y}}\bigl(y,y'\bigr) \bigr) \,d\nu(x,y) \,d\nu
\bigl(x',y'\bigr)
\\
& = & 4\int\int k_{\mathcal{X}}\bigl(x,x'\bigr)k_{\mathcal{Y}}
\bigl(y,y'\bigr) \,d\nu(x,y) \,d\nu\bigl(x',y'
\bigr)
\\
& = & 4\gamma_{k}^{2}(P_{XY},P_{X}P_{Y}),
\end{eqnarray*}
where we used that $\nu(\mathcal{X}\times\mathcal{Y})=0$, and that
$\int g(x,y,x',y') \,d\nu(x,y) \,d\nu(x',y')=0$ when $g$ does not
depend on one or more of its arguments, since $\nu$ also has zero
marginal measures. Convergence of integrals of the form
$\int k_{\mathcal{X}}(x,x)\times k_{\mathcal{Y}}(y,y) \,d\nu(x,y)$ is ensured
by the moment conditions on the marginals.
\end{pf}
We remark that a similar result to Theorem \ref{thmmdcovkern} is
given by \citeauthor{Lyons2011} [(\citeyear{Lyons2011}), Proposition 3.16], but without making
use of the link with kernel embeddings. Theorem~\ref{thmmdcovkern}
is a more general statement, in the sense that we allow $\rho$ to
be a semimetric of negative type, rather than metric. In addition,
the kernel interpretation leads to a significantly simpler proof:
the result is an immediate application of the HSIC expansion in \eqref
{eqHSICexpansion}.
%
\begin{rem}
As in Remark \ref{remfinitenessofmoments}, to ensure the existence
of the distance covariance, we impose a stronger condition on the
marginals: $P_{X}\in\mathcal{M}_{k_{\mathcal{X}}}^{2}(\mathcal{X})$
and $P_{Y}\in\mathcal{M}_{k_{\mathcal{Y}}}^{2}(\mathcal{Y})$, while
$P_{X}\in\mathcal{M}_{k_{\mathcal{X}}}^{1}(\mathcal{X})$ and
$P_{Y}\in
\mathcal{M}_{k_{\mathcal{Y}}}^{1}(\mathcal{Y})$
are sufficient for the existence of the Hilbert--Schmidt independence
criterion.
\end{rem}

By combining the Theorems \ref{thmm2sampledkern} and \ref{thmmdcovkern},
we can establish the direct relation between energy distance and distance
covariance, as discussed in Remark \ref{remdcov=00003Ddenergy?}.

\begin{cor}\label{cordcovandenergy}Let $(\mathcal{X},\rho
_{\mathcal{X}})$
and $(\mathcal{Y},\rho_{\mathcal{Y}})$ be semimetric spaces of negative
type, and let $X\sim P_{X}\in\mathcal{M}_{\rho_{\mathcal
{X}}}^{2}(\mathcal{X})$
and $Y\sim P_{Y}\in\mathcal{M}_{\rho_{\mathcal{Y}}}^{2}(\mathcal{Y})$,
having joint distribution $P_{XY}$. Then $\mathcal{V}_{\rho
_{\mathcal{X},}\rho_{\mathcal{Y}}}^{2}(X,Y)=D_{E,\tilde{\rho
}}(P_{XY},P_{X}P_{Y})$,
\textup{where} $\frac{1}{2}\tilde{\rho}$ is generated by the product
kernel in (\ref{eqtensorproductk}).
\end{cor}
%
\begin{rem}
As introduced by \citet{Szekely2007}, the notion of distance covariance
extends naturally to that of \emph{distance variance} $\mathcal
{V}^{2}(X)=\mathcal{V}^{2}(X,X)$
and of \emph{distance correlation} (by analogy with the Pearson product-moment
correlation coefficient),
\begin{eqnarray*}
\mathcal{R}^{2}(X,Y) & = & \cases{ \displaystyle\frac{\mathcal{V}^{2}(X,Y)}{\mathcal{V}(X)\mathcal{V}(Y)}, &\quad$
\mathcal{V}(X)\mathcal{V}(Y)>0,$\vspace*{2pt}
\cr
0, &\quad $\mathcal{V}(X)
\mathcal{V}(Y)=0.$}
\end{eqnarray*}
The distance correlation can also be expressed in terms of associated
kernels---see {Appendix}~\ref{secDistance-correlation} for
details.
\end{rem}

\subsection{Characteristic function
interpretation}
\label{subCharacteristic-function-interpre}

The distance covariance in \eqref{eqdCovintermsofdistances}
was defined by \citet{Szekely2007} in terms of a weighted distance
between characteristic functions. We briefly review this interpretation
here, and show that this approach \emph{cannot} be used to derive a
kernel-based measure of dependence [this result was first obtained
by \citet{GreFukSri09}, and is included here in the interest of completeness].
Let $X$ be a random vector on $\mathcal{X}=\mathbb{R}^{p}$ and $Y$
a random vector on $\mathcal{Y}=\mathbb{R}^{q}$. The characteristic
functions of $X$ and $Y$, respectively, will be denoted by $f_{X}$
and $f_{Y}$, and their joint characteristic function by $f_{XY}$.
The distance covariance $\mathcal{V}(X,Y)$ is defined via the norm
of $f_{XY}-f_{X}f_{Y}$ in a weighted $L_{2}$ space on $\mathbb{R}^{p+q}$,
that is,
%
%
\begin{equation}
\mathcal{V}^{2}(X,Y)=\int_{\mathbb{R}^{p+q}}\bigl\llvert
f_{X,Y}(t,s)-f_{X}(t)f_{Y}(s)\bigr\rrvert
^{2}w(t,s) \,dt \,ds\label{eqdcovviacharacteristic-1}
\end{equation}
for a particular choice of weight function given by
%
%
\begin{equation}
w(t,s)=\frac{1}{c_{p}c_{q}}\cdot\frac{1}{\llVert  t\rrVert
^{1+p}\llVert  s\rrVert ^{1+q}},\label{eqdcovweight-1}
\end{equation}
where $c_{d}=\pi^{{(1+d)}/{2}}/\Gamma({(1+d)}/{2})$, $d\geq1$.
An important property of distance covariance is that $\mathcal{V}(X,Y)=0$
if and only if $X$ and $Y$ are independent. We next obtain a similar
statistic in the kernel setting. Write $\mathcal{Z}=\mathcal{X}\times
\mathcal{Y}$,
and let $k(z,z')=\kappa(z-z')$ be a translation invariant RKHS kernel
on $\mathcal{Z}$, where $\kappa\dvtx \mathcal{Z}\to\mathbb{R}$ is a bounded
continuous function. Using Bochner's theorem, $\kappa$ can be written
as
\[
\kappa(z)  =  \int e^{-z^{\top}u}\,d\Lambda(u)
\]
for a finite nonnegative Borel measure $\Lambda$. It follows [\citet
{GreFukSri09}]
that
\[
\gamma_{k}^{2}(P_{XY},P_{X}P_{Y})=
\int_{\mathbb{R}^{p+q}}\bigl\llvert f_{X,Y}(t,s)-f_{X}(t)f_{Y}(s)
\bigr\rrvert ^{2} \,d\Lambda(t,s),
\]
which is in clear correspondence with \eqref{eqdcovviacharacteristic-1}.
The weight function in \eqref{eqdcovweight-1} is not integrable,
however, so we cannot find a continuous translation invariant kernel
for which $\gamma_{k}$ coincides with the distance covariance. Indeed,
the kernel in \eqref{eqtensorproductk} is \emph{not} translation
invariant.

A further related family of statistics for two-sample tests has been
studied by \citet{AlbaFernandez2008}, and the majority of results
therein can be directly obtained via Bochner's theorem from the corresponding
results on kernel two-sample testing, in the case of translation-invariant
kernels on $\mathbb{R}^{d}$. That being said, we emphasise that the
RKHS-based approach extends to general topological spaces and positive
definite functions, and it is unclear whether every kernel
two-sample/independence
test has an interpretation in terms of characteristic
functions.

\section{Distinguishing probability distributions}
\label{secDistinguishing-probability-distributions}

Theorem 3.20 of \citet{Lyons2011} shows that distance covariance
in a metric space characterizes independence if the metrics satisfy
an additional property, termed \emph{strong negative type}. We review
this notion and establish the interpretation of strong negative type
in terms of RKHS kernel properties.
%
\begin{defn}
The semimetric space $(\mathcal{Z},\rho)$, where $\rho$ is generated
by kernel $k$, is said to have \emph{a strong negative type} if
$\forall P,Q\in\mathcal{M}_{+}^{1}(\mathcal{Z})\cap\mathcal
{M}_{k}^{1}(\mathcal{Z})$,
%
%
\begin{equation}
P\neq Q\Rightarrow\int\rho \,d \bigl( [P-Q ]\times [P-Q ] \bigr)<0.\label{eqCISND}
\end{equation}
\end{defn}
Since the quantity in \eqref{eqCISND} is, by equation~\eqref{eqintegralexpressionenergydistance},
exactly $-D_{E,\rho}(P,Q)=-2\gamma_{k}^{2}(P,Q)$,
$\forall
P,Q\in\mathcal{M}_{+}^{1}(\mathcal{Z})\cap\mathcal
{M}_{k}^{1}(\mathcal{Z})$,
the following is immediate:
%
\begin{prop}
\label{prostrongnegativechar}Let kernel $k$ generate $\rho$.
Then $(\mathcal{Z},\rho)$ has a strong negative type if and only
if $k$ is characteristic to $\mathcal{M}_{+}^{1}(\mathcal{Z})\cap
\mathcal{M}_{k}^{1}(\mathcal{Z})$.
\end{prop}
Thus, the problem of checking whether a semimetric is of strong negative
type is equivalent to checking whether its associated kernel is characteristic
to an appropriate space of Borel probability measures. This conclusion
has some overlap with [\citet{Lyons2011}]: in particular, Proposition
\ref{prostrongnegativechar} is stated in [\citet{Lyons2011}, Proposition
3.10],
where the barycenter map $\beta$ is a kernel embedding in our terminology,
although Lyons does not consider distribution embeddings in an RKHS.
%
\begin{rem}
From \citet{Lyons2011}, Theorem 3.25, every separable Hilbert space
$\mathcal{Z}$ is of strong negative type, so a distance kernel $k$
induced by the (inner product) metric on $\mathcal{Z}$ is characteristic
to the appropriate space of probability measures.
\end{rem}

\begin{rem}
Consider the kernel in \eqref{eqtensorproductk}, and assume for
simplicity that $k_{\mathcal{X}}$ and $k_{\mathcal{Y}}$ are bounded,
so that we can consider embeddings of all probability measures. It
turns out that $k$ need not be characteristic---that is, it may not
be able to distinguish between any two distributions on $\mathcal
{X}\times\mathcal{Y}$,
even if $k_{\mathcal{X}}$ and $k_{\mathcal{Y}}$ are characteristic.
Namely, if $k_{\mathcal{X}}$ is the distance kernel induced by $\rho
_{\mathcal{X}}$
and centred at $x_{0}$, then $k((x_{0},y),(x_{0},y'))=0$ for all
$y,y'\in\mathcal{Y}$. That means that for every two distinct
$P_{Y},Q_{Y}\in\mathcal{M}_{+}^{1}(\mathcal{Y})$,
we have $\gamma_{k}^{2}(\delta_{x_{0}}P_{Y},\delta_{x_{0}}Q_{Y})=0$.
Thus, given that $\rho_{\mathcal{X}}$ and $\rho_{\mathcal{Y}}$ have
strong negative type, the kernel in \eqref{eqtensorproductk}
characterizes independence, but not equality of probability measures
on the product space. Informally speaking, distinguishing $P_{XY}$
from $P_{X}P_{Y}$ is an easier problem than two-sample testing on
the product space.
\end{rem}

\section{Empirical estimates and hypothesis tests}\label{secConsistency}

In this section, we outline the construction of tests based on the
empirical counterparts of MMD/energy distance and HSIC/distance
covariance.

\subsection{Two-sample testing}

So far, we have seen that the population expression of the MMD between
$P$ and $Q$ is well defined as long as $P$ and $Q$ lie in the
space $\mathcal{M}_{k}^{1/2}(\mathcal{Z})$, or, equivalently, have
a finite half-moment w.r.t. semimetric $\rho$ generated by $k$.
However, this assumption will not suffice to establish a meaningful
hypothesis test using empirical estimates of the MMD. We will require
a stronger condition, that $P,Q\in\mathcal{M}_{+}^{1}(\mathcal
{Z})\cap
\mathcal{M}_{k}^{1}(\mathcal{Z})$
(which is the same condition under which the energy distance is well
defined). Note that, under this condition we also have $k\in L_{P\times
P}^{2}(\mathcal{Z}\times\mathcal{Z})$,
as $\int\int k^{2}(z,z') \,dP(z) \,dP(z')\leq (\int k(z,z)
\,dP(z) )^{2}$.

Given i.i.d. samples $\mathbf{z}= \{ z_{i} \}
_{i=1}^{m}\sim P$
and $\mathbf{w}= \{ w_{i} \} _{i=1}^{n}\sim Q$, the empirical
(biased) $V$-statistic estimate of \eqref{eqMMD} is given by
%
%
\begin{eqnarray}\label{eqempiricalmmd}
\hat{\gamma}_{k,V}^{2}(\mathbf{z},\mathbf{w}) & = & \gamma
_{k}^{2} \Biggl(\frac{1}{m}\sum
_{i=1}^{m}\delta_{z_{i}},\frac{1}{n}
\sum_{j=1}^{n}\delta _{w_{j}} \Biggr)
\nonumber
\\
& = & \frac{1}{m^{2}}\sum_{i=1}^{m}\sum
_{j=1}^{m}k(z_{i},z_{j})+
\frac
{1}{n^{2}}\sum_{i=1}^{n}\sum
_{j=1}^{n}k(w_{i},w_{j})
\\
& &{} - \frac{2}{mn}\sum_{i=1}^{m}\sum
_{j=1}^{n}k(z_{i},w_{j}).
\nonumber
\end{eqnarray}
Recall that if $k$ generates $\rho$, this estimate involves only
the pairwise $\rho$-distances between the sample points.

We now describe a two-sample test using this statistic. The kernel
$\tilde{k}_{P}$ centred at $P$ in \eqref{eqcentredkernels} plays
a key role in characterizing the null distribution of degenerate $V$-statistic.
To $\tilde{k}_{P}$, we associate the \emph{integral kernel operator}
$S_{\tilde{k}_{P}}\dvtx L_{P}^{2}(\mathcal{Z})\to L_{P}^{2}(\mathcal{Z})$
[cf., e.g., \citet{Steinwart2008book}, page 126--127], given by
%
%
\begin{equation}
S_{\tilde{k}_{P}}g(z)  =  \int_{\mathcal{Z}}\tilde{k}_{P}(z,w)g(w)
\,dP(w).\label{eqkerneloperator}
\end{equation}
The condition that $P\in\mathcal{M}_{k}^{1}(\mathcal{Z})$, and, as
a consequence, that $\tilde{k}_{P}\in L_{P\times P}^{2}(\mathcal
{Z}\times\mathcal{Z})$,
is closely related to the desired properties of the integral operator.
Namely, this implies that $S_{\tilde{k}_{P}}$ is a trace class operator,
and, thus, a Hilbert--Schmidt operator [\citet{ReeSim80}, Proposition VI.23].
The following theorem is a special case of \citet{Gretton2012}, Theorem 12,
which extends \citet{AndHalTit94}, Section~2.3, to general RKHS kernels
(as noted by Anderson {et al.}, the form of the asymptotic distribution
of the $V$-statistic requires $S_{\tilde{k}_{P}}$ to be trace-class,
whereas the $U$-statistic has the weaker requirement that $S_{\tilde{k}_{P}}$
be Hilbert--Schmidt). For simplicity, we focus on the case where $m=n$.
%
\begin{thmm}
\label{thmmnull2sample}Let $k$ be a kernel on $\mathcal{Z}$,
and $\mathbf{\mathbf{Z}}= \{ Z_{i} \} _{i=1}^{m}$ and
$\mathbf
{W}= \{ W_{i} \} _{i=1}^{m}$
be two i.i.d. samples from $P\in\mathcal{M}_{+}^{1}(\mathcal{Z})\cap
\mathcal{M}_{k}^{1}(\mathcal{Z})$.
Assume $S_{\tilde{k}_{P}}$ is trace class. Then
%
%
\begin{equation}
\frac{m}{2}\hat{\gamma}_{k,V}^{2}(\mathbf{Z},
\mathbf{W})  \rightsquigarrow \sum_{i=1}^{\infty}
\lambda_{i}N_{i}^{2},\label{eqnulldist}
\end{equation}
where $N_{i}\stackrel{{i.i.d.}}{\sim}\mathcal{N}(0,1)$, $i\in\mathbb{N}$,
and $ \{ \lambda_{i} \} _{i=1}^{\infty}$ are the
eigenvalues of the operator~$S_{\tilde{k}_{P}}$.
\end{thmm}
Note that the limiting expression in \eqref{eqnulldist} is a valid
random variable precisely since $S_{\tilde{k}_{P}}$ is Hilbert--Schmidt,
that is, since $\sum_{i=1}^{\infty}\lambda_{i}^{2}<\infty$.

\subsection{Independence testing}
In the case of independence testing, we are given i.i.d. samples
$\mathbf{z}= \{ (x_{i},y_{i}) \} _{i=1}^{m}\sim P_{XY}$,
and the resulting $V$-statistic estimate (HSIC) is [\citeauthor{GreBouSmoSch05} (\citeyear{GreBouSmoSch05,GreFukTeoSonetal08})]
%
%
\begin{equation}
\operatorname{HSIC}(\mathbf{z};k_{\mathcal{X}},k_{\mathcal{Y}})=\frac
{1}{m^{2}}\operatorname{Tr}(K_{\mathcal{X}}HK_{\mathcal{Y}}H),
\end{equation}
where $K_{\mathcal{X}}$, $K_{\mathcal{Y}}$ and $H$ are $m\times m$
matrices given by $ (K_{\mathcal{X}} )_{ij}:=k_{\mathcal
{X}}(x_{i},x_{j})$,
$ (K_{\mathcal{Y}} )_{ij}:=k_{\mathcal{Y}}(y_{i},y_{j})$
and $H_{ij}=\delta_{ij}-\frac{1}{m}$ (centering matrix). The null
distribution of HSIC takes an analogous form to \eqref{eqnulldist}
of a weighted sum of chi-squares, but with coefficients corresponding
to the products of the eigenvalues of integral operators $S_{\tilde
{k}_{P_{X}}}\dvtx L_{P_{X}}^{2}(\mathcal{X})\to L_{P_{X}}^{2}(\mathcal{X})$
and $S_{\tilde{k}_{P_{Y}}}\dvtx L_{P_{Y}}^{2}(\mathcal{Y})\to
L_{P_{Y}}^{2}(\mathcal{Y})$.
Similarly to the case of two-sample testing, we will require that
$P_{X}\in\mathcal{M}_{k_{\mathcal{X}}}^{1}(\mathcal{X})$ and
$P_{Y}\in
\mathcal{M}_{k_{\mathcal{Y}}}^{1}(\mathcal{Y})$,
implying that integral operators $S_{\tilde{k}_{P_{X}}}$ and
$S_{\tilde
{k}_{P_{Y}}}$
are trace class operators. The following theorem is from \citet{Zhang2011},
Theorem 4.
See also \citet{Lyons2011}, Remark 2.9.
%
\begin{thmm}
\label{thmmnullindep-1}Let $\mathbf{\mathbf{Z}}= \{
(X_{i},Y_{i} ) \} _{i=1}^{m}$
be an i.i.d. sample from $P_{XY}=P_{X}P_{Y}$, with values in $\mathcal
{X}\times\mathcal{Y}$,
s.t. $P_{X}\in\mathcal{M}_{k_{\mathcal{X}}}^{1}(\mathcal{X})$ and
$P_{Y}\in\mathcal{M}_{k_{\mathcal{Y}}}^{1}(\mathcal{Y})$. Then
%
%
\begin{equation}
m\operatorname{HSIC}(\mathbf{Z};k_{\mathcal{X}},k_{\mathcal{Y}})  \rightsquigarrow\sum
_{i=1}^{\infty}\sum_{j=1}^{\infty}
\lambda_{i}\eta _{j}N_{i,j}^{2},
\label{eqnulldist-hsic-1}
\end{equation}
where $N_{i,j}\sim\mathcal{N}(0,1)$, $i,j\in\mathbb{N}$, are independent
and $ \{ \lambda_{i} \} _{i=1}^{\infty}$ and
$ \{
\eta_{j} \} _{j=1}^{\infty}$
are the eigenvalues of the operators $S_{\tilde{k}_{P_{X}}}$ and
$S_{\tilde{k}_{P_{Y}}}$, respectively.
\end{thmm}

\subsection{Test designs}
We would like to design distance-based tests with an asymptotic Type
I error of $\alpha$, and thus we require an estimate of the $
(1-\alpha )$-quantile
of the null distribution. We investigate two approaches, both of which
yield consistent tests: a bootstrap approach [\citet{ArcGin92}] and
a spectral approach [\citet{GreFukHarSri09,Zhang2011}]. The latter
requires empirical computation of eigenvalues of the integral kernel
operators, a problem studied extensively in the context of kernel
PCA [\citet{SchSmoMul97}]. To estimate limiting distribution in \eqref
{eqnulldist},
we compute the spectrum of the centred Gram matrix $\tilde{K}=HKH$
on the aggregated samples. Here, $K$ is a $2m\times2m$ matrix, with
entries $K_{ij}=k(u_{i},u_{j})$, $\mathbf{u}=[\mathbf{z} \mathbf{w}]$
is the concatenation of the two samples and $H$ is the centering
matrix. \citet{GreFukHarSri09} show that the null distribution defined
using the finite sample estimates of these eigenvalues converges to
the population distribution, provided that the spectrum is square-root
summable. As demonstrated empirically by \citet{GreFukHarSri09},
spectral estimation of the test threshold has a smaller computational
cost than that of the bootstrap-based approach, while providing
an indistinguishable performance. The same approach can be used in
obtaining a consistent finite sample null distribution for HSIC, via
computation of the empirical eigenvalues of $\tilde{K}_{\mathcal
{X}}=HK_{\mathcal{X}}H$
and $\tilde{K}_{\mathcal{Y}}=HK_{\mathcal{Y}}H$; see \citet{Zhang2011}.

Both \citeauthor{Szekely2004} [(\citeyear{Szekely2004}), page 14] and \citeauthor{Szekely2007} [(\citeyear{Szekely2007}), pages
2782--2783]
establish that the energy distance and distance covariance statistics,
respectively, converge to the weighted sums of chi-squares of forms
similar to \eqref{eqnulldist}. Analogous results for the generalized
distance covariance are presented in \citet{Lyons2011}, pages 7--8. These
works do not propose test designs that attempt to estimate the coefficients
$\lambda_{i}$, $i\in\mathbb{N}$, however. Besides the bootstrap,
\citeauthor{Szekely2007} [(\citeyear{Szekely2007}), Theorem 6] also propose an independence test
using a bound applicable to a general quadratic form $Q$ of centered
Gaussian random variables with $\mathbb{E}[Q]=1\dvtx\mathbb{P} \{
Q\geq (\Phi^{-1}(1-\alpha/2)^{2} ) \} \leq\alpha$,
valid for $0<\alpha\leq0.215$. When applied to the distance covariance
statistic, the upper bound of $\alpha$ is achieved if $X$ and $Y$
are independent Bernoulli variables. The authors remark that the resulting
criterion might be over-conservative. Thus, more sensitive distance
covariance tests are possible by computing the spectrum of the centred
Gram matrices associated to distance kernels, which is the approach
we apply in the next section.

\section{Experiments}\label{secExperiments}

In this section, we assess the numerical performance of the distance-based
and RKHS-based test statistics with some standard distance/kernel
choices on a series of synthetic data examples.

\subsection{Two-sample experiments}

\begin{figure}

\includegraphics{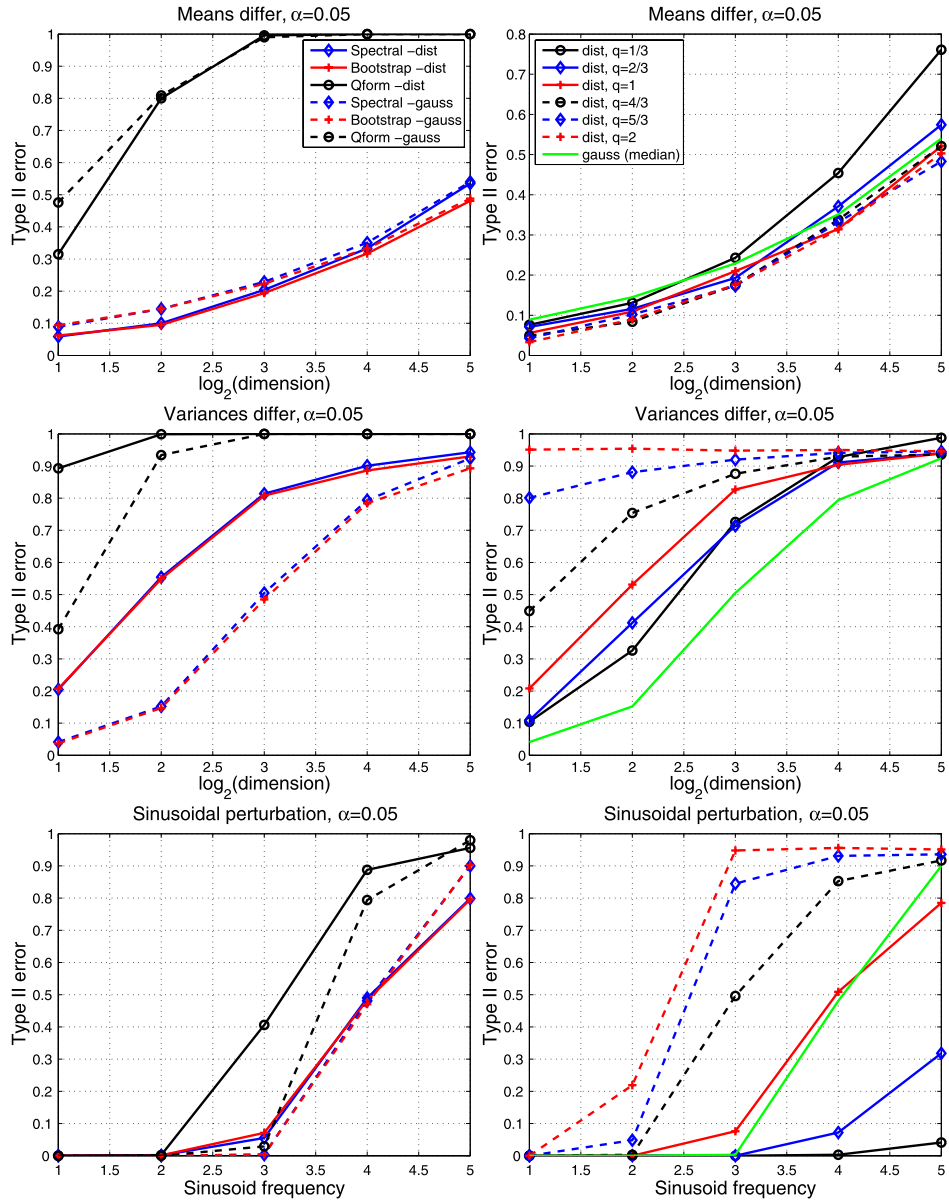}

\caption{(Left) MMD using Gaussian and distance
kernels for various tests; (right) Spectral MMD using distance kernels
with various exponents.}\label{figGaussian-vs-Dist}\vspace*{-3pt}
\end{figure}

In the two-sample experiments, we investigate three different kinds
of synthetic data. In the first, we compare two multivariate Gaussians,
where the means differ in one dimension only, and all variances are
equal. In the second, we again compare two multivariate Gaussians,
but this time with identical means in all dimensions, and variance
that differs in a single dimension. In our third experiment, we use
the benchmark data of \citet{SriFukGreLanetal09}: one distribution
is a univariate Gaussian, and the second is a univariate Gaussian
with a sinusoidal perturbation of increasing frequency (where higher
frequencies correspond to harder problems). All tests use a distance
kernel induced by the Euclidean distance. As shown on the left-hand
plots in Figure~\ref{figGaussian-vs-Dist}, the spectral and bootstrap
test designs appear indistinguishable, and significantly outperform
the test designed using the quadratic form bound, which appears to
be far too conservative for the data sets considered. The average
Type I errors are listed in Table~\ref{tabType-I-error}, and are
close to the desired test size of $\alpha=0.05$ for the spectral
and bootstrap tests.

\begin{table}
\tablewidth=200pt
\caption{Type I error (in \%) for two-sample tests
with distance-induced kernels}\label{tabType-I-error}
\begin{tabular*}{200pt}{@{\extracolsep{\fill}}lccc@{}}
\hline
& \textbf{mean} & \textbf{var} & \multicolumn{1}{c@{}}{\textbf{sine}}\\
\hline
Spec & 4.66 & 4.72 & 5.10\\
Boot & 5.02 & 5.16 & 5.20\\
Qform & 0.02 & 0.05 & 0.98\\
\hline
\end{tabular*}
\end{table}

We also compare the performance to that of the Gaussian kernel, commonly
used in machine learning, with the bandwidth set to the median distance
between points in the aggregation of samples. We see that when the
means differ, both tests perform similarly. When the variances differ,
it is clear that the Gaussian kernel has a major advantage over the
distance-induced kernel, although this advantage decreases with increasing
dimension (where both perform poorly). In the case of a sinusoidal
perturbation, the performance is again very similar.

In addition, following Example \ref{exavariousexponents}, we investigate
performance of kernels obtained using the semimetric $\rho(z,z')=\llVert  z-z'\rrVert ^{q}$
for $0<q\leq2$. Results are presented in the right-hand plots of
Figure~\ref{figGaussian-vs-Dist}. In the case of sinusoidal perturbation,
we observe a dramatic improvement compared with the $q=1$ case and
the Gaussian kernel: values $q=1/3$ (and smaller) offer virtually
error-free performance even at high frequencies [note that $q=1$
yields the energy distance described in \citeauthor{Szekely2004} (\citeyear{Szekely2004,Szekely2005})].
Small improvements over a wider $q$ range are also observed in the
cases of differing mean and variance.

We observe from the simulation results that distance-induced kernels
with higher exponents are advantageous in cases where distributions
differ in mean value along a single dimension (with noise in the remainder),
whereas distance kernels with smaller exponents are more sensitive
to differences in distributions at finer lengthscales (i.e., where
the characteristic functions of the distributions differ at higher
frequencies).

\subsection{Independence experiments}
\begin{figure}

\includegraphics{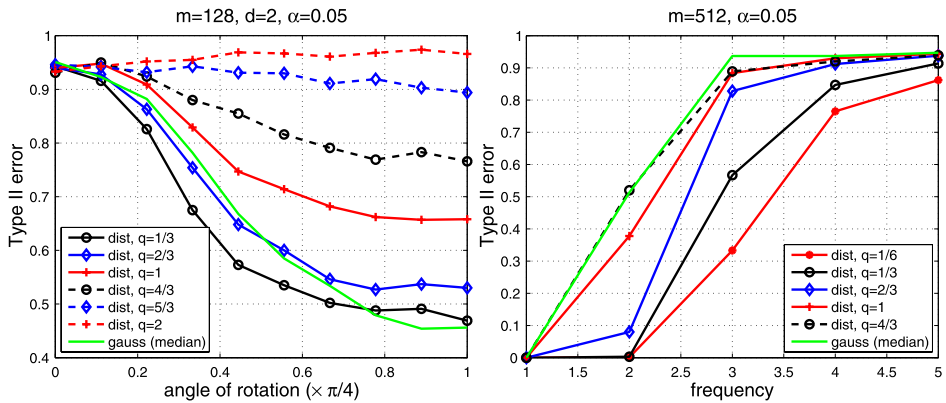}

\caption{HSIC using distance kernels with various
exponents and a Gaussian kernel as a function of (left) the angle
of rotation for the dependence induced by rotation; (right) frequency
$\ell$ in the sinusoidal dependence example.}\label{figqdistHSICALL}
\end{figure}

To assess independence tests, we used an artificial benchmark proposed
by \citet{GreFukTeoSonetal08}: we generated univariate random variables
from the Independent Component Analysis (ICA) benchmark densities
of \citet{BacJor02}; rotated them in the product space by an angle
between $0$ and $\pi/4$ to introduce dependence; filled additional
dimensions with independent Gaussian noise; and, finally, passed the
resulting multivariate data through random and independent orthogonal
transformations. The resulting random variables $X$ and $Y$ were
dependent but uncorrelated. The case $m=128$ (sample size) and $d=2$
(dimension) is plotted in Figure~\ref{figqdistHSICALL} (left).
As observed by \citet{GreFukSri09}, the Gaussian kernel using the
median inter-point distance as bandwidth does better than the distance-induced
kernel with $q=1$. By varying $q$, however, we are able to obtain
a wide performance range: in particular, the values $q=1/3$ (and
smaller) have an advantage over the Gaussian kernel on this dataset.
As for\vadjust{\goodbreak} the two-sample case, bootstrap and spectral tests have indistinguishable
performance, and are significantly more sensitive than the quadratic
form-based test, which failed to detect any dependence on this dataset.

%
%
%
%
%

In addition, we assess the performance on sinusoidally dependent data.
The sample of the random variable pair $X,Y$ was drawn from
$P_{XY}\propto1+\sin(\ell x)\times \sin(\ell y)$
for integer $\ell$, on the support $\mathcal{X}\times\mathcal{Y}$,
where $\mathcal{X}:=[-\pi,\pi]$ and $\mathcal{Y}:=[-\pi,\pi]$. In
this way, increasing $\ell$ causes the departure from a uniform (independent)
distribution to occur at increasing frequencies, making this departure
harder to detect given a small sample size. Results are in Figure~\ref{figqdistHSICALL} (right). The distance covariance outperforms
the Gaussian kernel (median bandwidth) on this example, and smaller
exponents result in better performance (lower Type II error when the
departure from independence occurs at higher frequencies). Finally,
we note that the setting $q=1$, as described by \citet{Szekely2007,SzeRiz09},
is a reasonable heuristic in practice, but does not yield the most
powerful tests on either dataset. Informally, the exponent in the
distance-induced kernel plays a similar role as the bandwidth of the
Gaussian kernel, and smaller exponents are able to detect dependencies
at smaller lengthscales. Poor performance of the Gaussian kernel with
median bandwidth in this example is a consequence of the mismatch
between the overall lengthscale of the marginal distributions (captured
by the median inter-point distances) and the lengthscales at which
dependencies are present.

%
%

\section{Conclusion}

We have established an equivalence between the generalized notions
of energy distance and distance covariance, computed with respect
to semimetrics of negative type, and distances between embeddings
of probability measures\vadjust{\goodbreak} into certain reproducing kernel Hilbert spaces.
As a consequence, we can view energy distance and distance covariance
as members of a much larger class of discrepancy/dependence measures,
and we can choose among this larger class to design more powerful
tests. For instance, \citet{Gretton2012a} recently proposed a strategy
of selecting from a candidate kernels so as to asymptotically optimize
the relative efficiency of a two-sample test. Moreover, kernel-based
tests can be performed on the data that do not lie in a Euclidean
space. This opens the door to new and powerful tools for exploratory
data analysis whenever an appropriate domain-specific notion of distance
(negative type semimetric) or similarity (kernel) can be defined.
Finally, the family of kernels that arises from the energy distance/distance
covariance can be employed in many additional kernel-based applications
in statistics and machine learning, such as conditional dependence
testing and estimating the chi-squared distance [\citet{FukGreSunSch08}],
Bayesian inference [\citet{NIPS20110985}] and mixture density estimation
[\citet{Sriperumbudur11}].

\begin{appendix}
\section{Distance correlation}
\label{secDistance-correlation}
As described by \citet{Szekely2007},
the notion of distance covariance extends naturally to that of \emph{distance
variance} $\mathcal{V}^{2}(X)=\mathcal{V}^{2}(X,X)$ and of \emph{distance
correlation} (by analogy with the Pearson product-moment correlation
coefficient),
\[
\mathcal{R}^{2}(X,Y) =  \cases{\displaystyle \frac{\mathcal{V}^{2}(X,Y)}{\mathcal{V}(X)\mathcal{V}(Y)}, &\quad $
\mathcal{V}(X)\mathcal{V}(Y)>0,$\vspace*{2pt}
\cr
0, &\quad $\mathcal{V}(X)
\mathcal{V}(Y)=0.$}
\]
Distance correlation also has a straightforward interpretation in
terms of kernels,
%
%
\begin{eqnarray}\label{eqdcor}
\mathcal{R}^{2}(X,Y) & = & \frac{\mathcal{V}^{2}(X,Y)}{\mathcal
{V}(X)\mathcal{V}(Y)}
\nonumber
\\
& = & \frac{\gamma_{k}^{2}(P_{XY},P_{X}P_{Y})}{\gamma
_{k}(P_{XX},P_{X}P_{X})\gamma_{k}(P_{YY},P_{Y}P_{Y})}
\\
& = & \frac{\llVert \Sigma_{XY}\rrVert _{\mathrm{HS}}^{2}}{\llVert
\Sigma_{XX}\rrVert _{\mathrm{HS}}\llVert \Sigma_{YY}\rrVert
_{\mathrm{HS}}},\nonumber
\end{eqnarray}
where covariance operator $\Sigma_{XY}\dvtx \mathcal{H}_{k_{\mathcal
{X}}}\to
\mathcal{H}_{k_{\mathcal{Y}}}$
is a linear operator for which $ \langle\Sigma_{XY}f,g
\rangle_{\mathcal{H}_{k_{\mathcal{Y}}}}=\mathbb{E}_{XY}
[f(X)g(Y) ]-\mathbb{E}_{X}f(X)\mathbb{E}_{Y}g(Y)$
for all $f\in\mathcal{H}_{k_{\mathcal{X}}}$ and $g\in\mathcal
{H}_{k_{\mathcal{Y}}}$,
and $\llVert \cdot\rrVert _{\mathrm{HS}}$ denotes the Hilbert--Schmidt
norm [\citet{GreBouSmoSch05}]. It is clear that $\mathcal{R}$ is invariant
to scaling $(X,Y)\mapsto(\varepsilon X,\varepsilon Y)$, $\varepsilon>0$,
whenever the corresponding semimetrics are homogeneous, that is, whenever
$\rho_{\mathcal{X}}(\varepsilon x,\varepsilon x')=\varepsilon\rho_{\mathcal
{X}}(x,x')$,
and similarly for $\rho_{\mathcal{Y}}$. Moreover, $\mathcal{R}$
is invariant to translations, $(X,Y)\mapsto(X+x',Y+y')$, $x'\in
\mathcal{X}$,
$y'\in\mathcal{Y}$, whenever $\rho_{\mathcal{X}}$ and $\rho
_{\mathcal{Y}}$
are translation invariant. Therefore, by varying the choice of kernels
$k_{\mathcal{X}}$ and $k_{\mathcal{Y}}$, we obtain in \eqref{eqdcor}
a very broad class of dependence measures that generalize the distance
correlation of \citet{Szekely2007} and can be used in exploratory
data analysis as a measure of dependence between pairs of random variables
that take values in multivariate or structured/non-Euclidean domains.

\section{Link with universal kernels}
\label{secLink-universal-kernels}
We briefly remark on how our results
on equivalent kernels relate to the notion of universal kernels on
compact metric spaces in the sense of \citet{Steinwart2008book}, Definition
4.52:
%
\begin{defn}
A continuous kernel $k$ on a compact metric space $\mathcal{Z}$
is said to be universal if its RKHS $\mathcal{H}_{k}$ is dense in
the space $C(\mathcal{Z})$ of continuous functions on $\mathcal{Z}$,
endowed with the uniform norm.
\end{defn}
The family of universal kernels includes the most
popular choices in machine learning literature, including the Gaussian
and the Laplacian kernel. The following characterization of universal
kernels is due to \citet{Sriperumbudur2011}:
%
\begin{prop}
Let $k$ be a continuous kernel on a compact metric space $\mathcal{Z}$.
Then, $k$ is universal if and only if $\mu_{k}\dvtx \mathcal{M}(\mathcal
{Z})\to\mathcal{H}_{k}$
is a vector space monomorphism, that is,
\[
\bigl\llVert \mu_{k}(\nu)\bigr\rrVert _{\mathcal{H}_{k}}^{2}=
\int \int k\bigl(z,z'\bigr) \,d\nu(z) \,d\nu\bigl(z'\bigr)>0\qquad
\forall\nu\in\mathcal{M}(\mathcal {Z})\setminus\{0\}.
\]
\end{prop}
As a direct consequence, every universal kernel $k$
is also characteristic, as $\mu_{k}$ is, in particular, injective
on the space of probability measures. Now, consider a kernel $\tilde{k}_{f}$
centered at $f=\mu_{k}(\nu)$ for some $\nu\in\mathcal{M}(\mathcal{Z})$,
such that $\nu(\mathcal{Z})=1$. Then $\tilde{k}_{f}$ is no longer
universal, since
\begin{eqnarray*}
&&\bigl\llVert \mu_{\tilde{k}_{f}}(\nu)\bigr\rrVert _{\mathcal
{H}_{\tilde
{k}_{f}}}^{2}
\\
&&\qquad =  \int\tilde{k}_{f}\bigl(z,z'\bigr) \,d\nu(z) \,d\nu
\bigl(z'\bigr)
\\
&&\qquad =  \int\int \biggl[k\bigl(z,z'\bigr)-\int k(w,z) \,d\nu(w)-\int k
\bigl(w,z'\bigr) \,d\nu (w)
\\
& &\qquad\hspace*{108pt}{} +\int\int k\bigl(w,w'\bigr) \,d\nu(w) \,d\nu\bigl(w'
\bigr) \biggr] \,d\nu(z) \,d\nu\bigl(z'\bigr)
\\
&&\qquad =  \bigl(1-\nu(\mathcal{Z}) \bigr)^{2}\bigl\llVert
\mu_{k}(\nu )\bigr\rrVert _{\mathcal{H}_{k}}^{2}
\\
&&\qquad =  0.
\end{eqnarray*}
However, $\tilde{k}_{f}$ is still characteristic, as it is equivalent
to $k$. This means that all kernels of the form \eqref{eqcentredkernels},
including the distance kernels, are examples of nonuniversal characteristic
kernels, provided that they generate a semimetric $\rho$ of strong
negative type. In particular, the kernel in (\ref{eqeuclideankernel})
on a compact $\mathcal{Z}\subset\mathbb{R}^{d}$ is a characteristic
nonuniversal kernel for $q<2$. This result is of some interest to
the machine learning community, as such kernels have typically been
difficult to construct. For example, the two notions are known to
be equivalent on the family of translation invariant kernels on
$\mathbb{R}^{d}$
[\citet{Sriperumbudur2011}].
\end{appendix}

\section*{Acknowledgments}
D.~Sejdinovic, B.~Sriperumbudur and A.~Gretton acknowledge support of the
Gatsby Charitable Foundation. The work was carried out when
B.~Sriperumbudur was with Gatsby Unit, University College London.
B.~Sriperumbudur and A.~Gretton contributed equally.

%


\printaddresses

\end{document}